\documentclass[
 reprint,
nofootinbib,
 amsmath,amssymb,
 aps,
]{revtex4-2}
\usepackage[nolist]{acronym} 
\usepackage{siunitx} 
\usepackage{physics}
\usepackage{multirow}
\usepackage{amsmath}
\usepackage[version=4]{mhchem}
\usepackage{graphicx}% Include figure files
\usepackage{dcolumn}% Align table columns on decimal point
\usepackage{bm}% bold math
\usepackage{color}
\usepackage{textcomp}
\usepackage{hyperref}% add hypertext capabilities
\newcommand{\diff}{\mathrm{d}}

\usepackage[normalem]{ulem}

\begin{document}

\preprint{APS/123-QED}

\title{Detecting the diffuse supernova neutrino background in the future water-based liquid scintillator detector Theia}

\author{Julia Sawatzki}
 \email{julia.sawatzki@ph.tum.de}
 \affiliation{Physik-Department, Technische Universit\"at M\"unchen, James-Franck-Stra\ss e 1, 85748 Garching, Germany}

\author{Michael Wurm}
 \email{michael.wurm@uni-mainz.de}
\affiliation{Institut f\"ur Physik, Johannes Gutenberg Universität Mainz, Staudinger Weg 7, 
55128 Mainz, Germany}
\author{Daniel Kresse}
\email{danielkr@mpa-garching.mpg.de}
\affiliation{Max-Planck-Institut f\"ur Astrophysik, Karl-Schwarzschild-Stra\ss e~1, 85748 Garching, Germany}
\affiliation{Physik-Department, Technische Universit\"at M\"unchen, James-Franck-Stra\ss e 1, 85748 Garching, Germany}
\date{21 January 2021}

\begin{abstract}
A large-scale neutrino observatory based on Water-based Liquid Scintillator (WbLS) will be excellently suited for a measurement of the Diffuse Supernova Neutrino Background (DSNB). The WbLS technique offers high signal efficiency and effective suppression of the otherwise overwhelming background from neutral-current interactions of atmospheric neutrinos. To illustrate this, we investigate the DSNB sensitivity for two configurations of the future Theia detector by developing the expected signal and background rejection efficiencies along a full analysis chain. Based on a statistical analysis of the remaining signal and background rates, we find that a rather moderate exposure of 190\,kt$\cdot$yrs will be sufficient to claim a ($5\sigma$) discovery of the faint DSNB signal for standard model assumptions. We conclude that, in comparison with other experimental techniques, WbLS offers the highest signal efficiency of more than 80\,\% and best signal significance over background.
\end{abstract}

%Use showkeys class option if keyword
                              %display desired
\maketitle

%\tableofcontents
\begin{acronym}[Bash]
\acro{SK}[SK]{Super-Kamiokande}
\acro{DUNE}{Deep Underground Neutrino Experiment} 
\acro{NC}{neutral-current}
\acro{C/S}{Cherenkov-to-Scintillation}
\acro{DSNB}[DSNB]{Diffuse Supernova Neutrino Background}
\end{acronym}

%----- INTRODUCTION --------
\section{\label{sec:introduction}Introduction}
Core-collapse Supernovae (SNe) are intense sources of low-energy neutrinos ($E_\nu\lesssim 50$\,MeV). For SNe occurring within the Milky Way, bright signals are expected for several current-day neutrino detectors, permitting detailed analyses of the astrophysics of the explosion and superimposed effects caused by neutrino properties  (for a comprehensive review, see e.g. Ref.\ \cite{Mirizzi:2015eza}). However, even compared to the decades-long operation times of large-volume neutrino observatories, galactic SNe are rare. Therefore, the search for the faint but constant signal predicted for the Diffuse Supernova Neutrino Background (DSNB), i.e.\ the integral neutrino flux from past core-collapse SNe at cosmological distances, is especially appealing \cite{Ando:2004hc,Beacom:2010kk,Vissani:2011kx,Lunardini:2012ne,Nakazato:2015rya,Horiuchi:2017qja,Priya:2017bmm,Moller:2018kpn,Riya:2020wpw,Kresse:2020}.\ A measurement of the DSNB flux and spectrum will provide valuable information on the redshift-dependent SN rate as well as on the properties of stellar core collapse, such as the nuclear equation of state of the emerging neutron stars or the fraction of core-collapse events leading to the formation of black holes (BHs) in faint or failed explosions.

Given the minute expected flux of ${\cal O}(10^2)$ per cm$^2$s and low energy of DSNB neutrinos and anti\-neutrinos of all flavors, an experimental measurement is very challenging. Detector target masses on the order of $\sim$10 kilotons are required to obtain one signal event per year.\ Today, the \ac{SK} experiment holds the best upper limit on the DSNB's $\bar\nu_e$ flux component of (2.8$-$3.1)\,\si{cm^{-2} s^{-1}} above \SI{17.3}{MeV} \cite{Bays:2011si}.\ At the time of writing, the \ac{SK} collaboration prepares an upgrade to the detector by dissolving gadolinium salt in the water target to greatly enhance neutron detection capabilities \cite{Sekiya:2016xji}. In parallel, the JUNO experiment in southern China is entering its construction phase with the first data expected two years from now \cite{An:2015jdp}. Both experiments are likely to provide the first evidence ($3\sigma$) of the DSNB signal. However, given that the fiducial target masses of $\sim$20\,kt are relatively low by standards of the DSNB, accumulation of event statistics will be slow. Moreover, the presence of background events caused by neutral-current (NC) interactions of atmospheric neutrinos complicates detection~\cite{Moller:2018kpn,An:2015jdp,Collaboration:2011jza}.

The present paper studies the potential of an advanced detection concept for a definitive $5\sigma$-detection of a "standard" DSNB signal \cite{Kresse:2020}. As has been laid out in the white paper of the future Theia detector \cite{Askins:2020aa}, Water-based Liquid Scintillator (WbLS) in combination with ultra-fast light sensors (LAPPDs) and/or high PMT granularity permits the simultaneous detection of Cherenkov and scintillation light. %%
Different from pure water or organic scintillator detectors, the evaluation of the dual Cherenkov/scintillation signal provides superior background discrimination. Correspondingly, a WbLS detector will feature an excellent signal-to-background ratio: For Theia at Homestake, we expect $\sim$ $17_{-8}
^{+34}$ DSNB signal events over $9\pm 2$ background events for 100\,kt$\cdot$yrs~(sec.\,\ref{sec:sensitivity}) and an observation window ranging from 8 to 30\,MeV.

Note that a similar study has been performed in Ref.~\cite{Wei:2016vjd} in the context of the Jinping Neutrino Experiment.\ However, the present paper goes substantially beyond the earlier study by including a realistic detector simulation to evaluate the expected ratio of Cherenkov and scintillation photons (C/S ratio) detected.\ Moreover, we investigate not only background discrimination based on the ratio of detected Cherenkov and scintillation photons but as well the possibility to add further background tags exploiting Cherenkov ring counting and delayed decays of excited final-state nuclei.

The paper is structured as follows: Sec.\,\ref{sec:wbls} sets out the basic concept of DSNB detection in WbLS and the layout of the Theia detectors. 
The recent DSNB flux models by \cite{Kresse:2020} employed in this work are briefly described in sec.\,\ref{sec:dsnb}. Relevant backgrounds and several conventional techniques for their suppression in liquid-scintillator detectors are shortly reviewed in sec.\,\ref{sec:background}. Instead, sec.\,\ref{sec:bg_reduction} places particular emphasis on the discrimination techniques specific to WbLS that are instrumental to the very effective reduction of the dominant NC background from atmospheric neutrinos: delayed decays from oxygen spallation, ring counting, and $-$ most importantly $-$ the C/S ratio. Sec.\,\ref{sec:sensitivity} provides signal and background rates as well as the corresponding DSNB discovery potential, while sec.\,\ref{sec:others} puts these results into context with other existing and planned neutrino detectors.

% -----WBLS --------
\section{\label{sec:wbls}Detector Technology}
 When neutrinos interact in a conventional (i.e.\ organic) liquid scintillator detector, the final state particles will create not only scintillation but also Cherenkov photons. However, both organic solvent and fluorophores feature strong absorption bands in the dominant UV/blue part of the Cherenkov spectrum. Upon arrival at the photosensors surrounding the scintillator neutrino target, the remaining direct (i.e.\ unscattered) Cherenkov photons are effectively hard to distinguish from the overwhelming  scintillation signal.
 \medskip\\
 {\bf Water-based Liquid Scintillator (WbLS).} The primary motivation for the use of WbLS is to provide a very transparent scintillator by adding ultrapure water to the organic compounds.\ The resulting liquid is an emulsion that consists of mycels (nanoscopic droplets) of organic material surrounded by a surfactant and dissolved in the bulk water.\ The main characteristics of state-of-the-art WbLS samples have been investigated in \cite{Bignell:2015oqa,Caravaca:2020lfs}: The light yield of the resulting WbLS is roughly proportional to its organic fraction.\ The attenuation length reached in the blue spectral range depends primarily on the properties of the diluted organic fraction and is assumed here very conservatively to $\sim$20\,m \cite{Askins:2020aa}. This permits the detection of a sizable fraction of the Cherenkov photons.
 \medskip\\
{\bf Cherenkov/scintillation separation.} To make use of the information carried by the Cherenkov signal, it must be distinguished from the scintillation photons \cite{Land:2020oiz}. For this, there are three basic approaches:
\begin{itemize}
    \item fast light sensors with (sub-)nanosecond time resolution (e.g.\ LAPPDs \cite{Lyashenko:2019tdj}) permit the identification of a front of Cherenkov light arriving nanoseconds earlier than the delayed scintillation emission, 
    \item the characteristic angular dependence of Cherenkov emission will cause a ring-shaped local enhancement in the detected light intensity on top of the isotropic scintillation signal, and
    \item wavelength-sensitive photosensors, e.g.\ dichroi\-cons~\cite{Kaptanoglu:2019gtg},
    can distinguish near-UV scintillation light from the blue-green lower end of the Cherenkov spectrum that has been traveling unperturbedly through the WbLS bulk volume.
\end{itemize}

\begin{figure}[tb]
\centering
\includegraphics[width=0.5\textwidth]{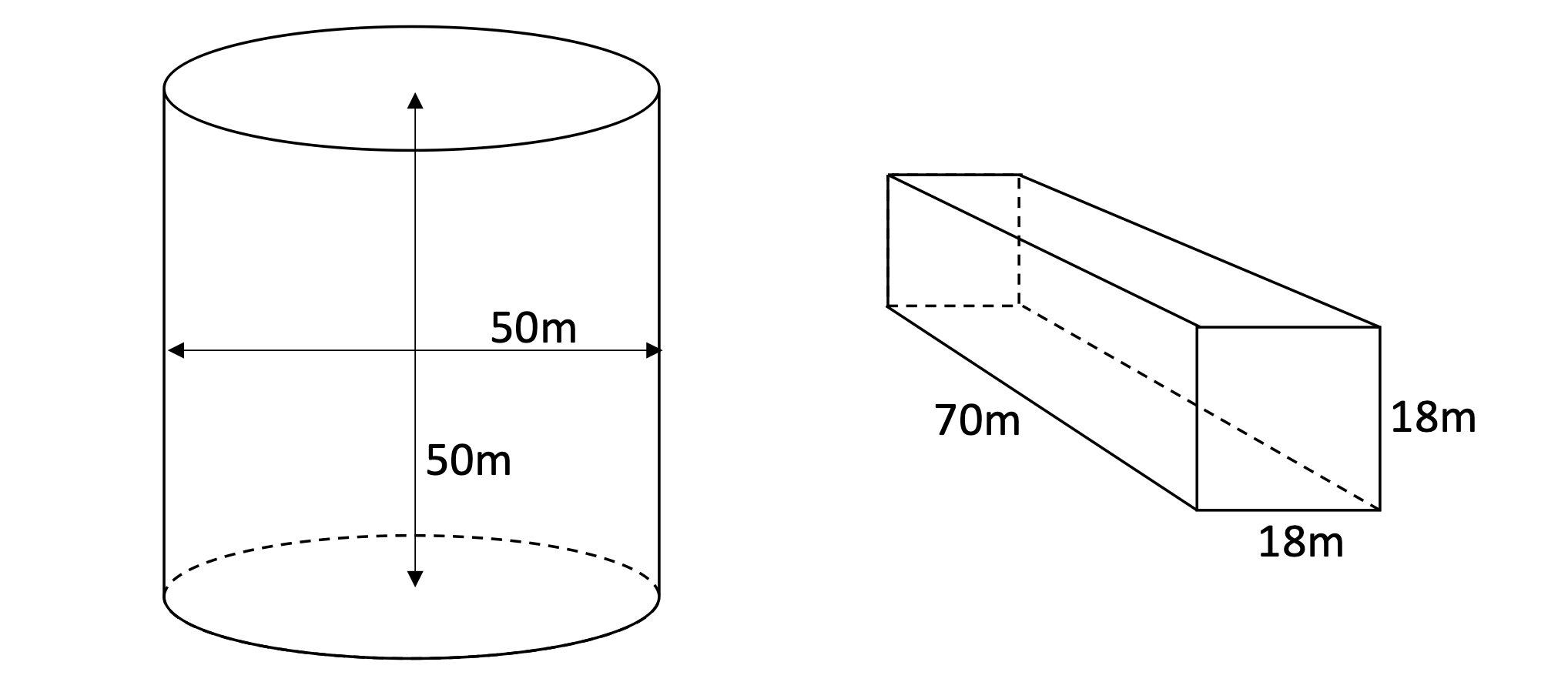}
\caption{\label{fig:theia}Basic detector geometries of Theia100 (left) and Theia25 (right).}
\end{figure}

\noindent {\bf Theia.} A large-volume WbLS detector has been first proposed as part of the Advanced Scintillator Detection Concept \cite{Alonso:2014fwf}.\ It has been later on developed further to the Theia detector project \cite{Askins:2020aa}: The full physics potential of the WbLS technique could be exploited with a detector mass of 100\,kt, named Theia100.\ More recently, a smaller detector realization, Theia25, has been discussed as a possible contender for the fourth DUNE detector module with a WbLS mass of 25\,kt \cite{Askins:2020aa}. The corresponding detector geometries are displayed in fig.\,\ref{fig:theia}. Here, we consider a deployment at the 4850 foot level in Homestake/SURF laboratory (4300\,m\,w.\,e.).
\medskip\\
{\bf DSNB detection.} The detection potential of the DSNB in WbLS has been first described in \cite{Alonso:2014fwf}.\ As will be demonstrated in sec.\,\ref{sec:bg_reduction}, the possibility to detect and distinguish Cherenkov and scintillation photons provides a great advantage in terms of background discrimination when compared to conventional water Cherenkov or fully organic scintillation detectors: Large-volume Cherenkov detectors like \ac{SK} I-IV feature low tagging efficiency for the delayed neutron capture in inverse beta decays (IBDs), making event selection much more susceptible to single-event backgrounds \cite{Malek:2002ns,Watanabe:2008ru,Bays:2011si,Zhang:2013tua}.\ On the other hand, liquid scintillator detectors like JUNO suffer from neutral-current (NC) interactions of atmospheric neutrinos on carbon that may break-up the nucleus and create IBD-like signatures from quenched final-state hadrons and the delayed capture of a single neutron
~\cite{Wurm:2007cy,Wurm:2011zn,Collaboration:2011jza,Mollenberg:2014pwa,An:2015jdp}. A similar background from NC reactions on oxygen is present as well in \ac{SK} \cite{Wan:2019xnl}.\ It should be noted that both detector technologies can mitigate those backgrounds: \ac{SK} will soon start a detection run with gadolinium added to the water, named SK-Gd, greatly enhancing the delayed neutron tag \cite{Beacom:2003nk}.\ On the other hand, pulse-shape discrimination may be used in organic scintillators to suppress the NC background \cite{Mollenberg:2014pwa, An:2015jdp}. 
\medskip\\
{\bf DSNB in Theia.} For both water Cherenkov and organic scintillator experiments, the residual detection efficiency for the DSNB signal is in the order of 50\,\%. Contrariwise, we expect a detection efficiency of more than 80\,\% in WbLS. As in an organic scintillator, the additional scintillation light enables a reliant detection of the delayed neutron capture.\ Crucially, background discrimination capabilities are greatly enhanced by exploiting Cherenkov ring counting (sec.\,\ref{sec:ringcounting}) and the Cherenkov-scintillation ratio (sec.\,\ref{sec:csratio}). As a result, the expected signal-to-background ratio will be excellent. Moreover, a detailed study of NC atmospheric neutrino background events with the diagnostic possibilities of WbLS detection is likely to provide SK-Gd and JUNO with a better systematic understanding of their respective NC background levels and event topologies.
\medskip\\
{\bf Detector configuration.} In the Geant4 (Version~9.4.p4) simulation underlying this study \cite{Agostinelli:2002hh,Allison:2006ve}, we used a generic spherical detector geometry on the scale of Theia25. We assume a WbLS with a 10\,\% organic fraction, corresponding to a scintillation light yield of $\sim$\num{d3} photons per MeV and thus on the same scale as the Cherenkov light emission. Light transport and scattering in the target medium have been implemented. The absorption and Rayleigh scattering length were set to \SI{77}{m} and \SI{27}{m}, respectively, to obtain an attenuation length of $\sim$\,\SI{20}{m} (for $\lambda=$\,\SI{430}{nm}). Since data on quenching in WbLS is scarce \cite{Bignell:2015oqa}, and none is available for the relevant energy range, we assumed the quenching factors to be similar to that of organic scintillators.

Furthermore, dense instrumentation of the detector surfaces is required to obtain both sizable Cherenkov and scintillation signals. In Theia, this will be realized by a mixture of high-QE 10''-PMTs and LAPPDs (ratio 25:1). Here, we assume a generic coverage of 70\,\%~(25\,\%) of the detector surface for Theia100 (Theia25) and a peak detection efficiency of $\sim$\,30\% for the PMT-like photosensors~\cite{Askins:2020aa}. The resulting photoelectron yield for the scintillation signal is 130\,p.e./MeV, while the Cherenkov component provides 80\,p.e./MeV.\ Note that, in the following, all visible energy spectra are based solely on the photons collected for the scintillation component of the signal.\ This implicitly assumes close-to-perfect separation of the scintillation and Cherenkov components. The same assumption is made for the event selection in sec.\,\ref{sec:bg_reduction}. 
\medskip\\ \newpage
\textbf{Inverse beta decay.} Since it has the largest cross-section at low energies, the primary detection channel for the DSNB in water and organic liquids is the inverse beta decay (IBD) of electron antineutrinos on free protons: $\bar{\nu}_e + p \rightarrow e^+ + n$.
Due to the kinematics of the IBD reaction, the kinetic energy of the positron translates almost directly to the incident neutrino energy but is reduced by the reaction Q-value of \SI{1.8}{MeV} \cite{Strumia:2003zx}. Hence, the neutrino energy can be reconstructed from the detected visible energy ($E_{\nu} \approx E_{\rm vis}+ \SI{0.8}{MeV}$), considering that the final-state positron annihilates with an electron into two \SI{511}{keV} photons. The produced neutron gets captured by a proton with a mean capture time of $\sim$250\,\textmu s, producing a \SI{2.2}{MeV} gamma-ray well visible in liquid scintillator. Hence, the delayed neutron signal provides a fast coincidence tag to reduce the ample single-event backgrounds.

% -----DSNB MODEL --------
\section{\label{sec:dsnb}DSNB Model}
\begin{figure}[tb]
\centering
\includegraphics[width=0.5\textwidth]{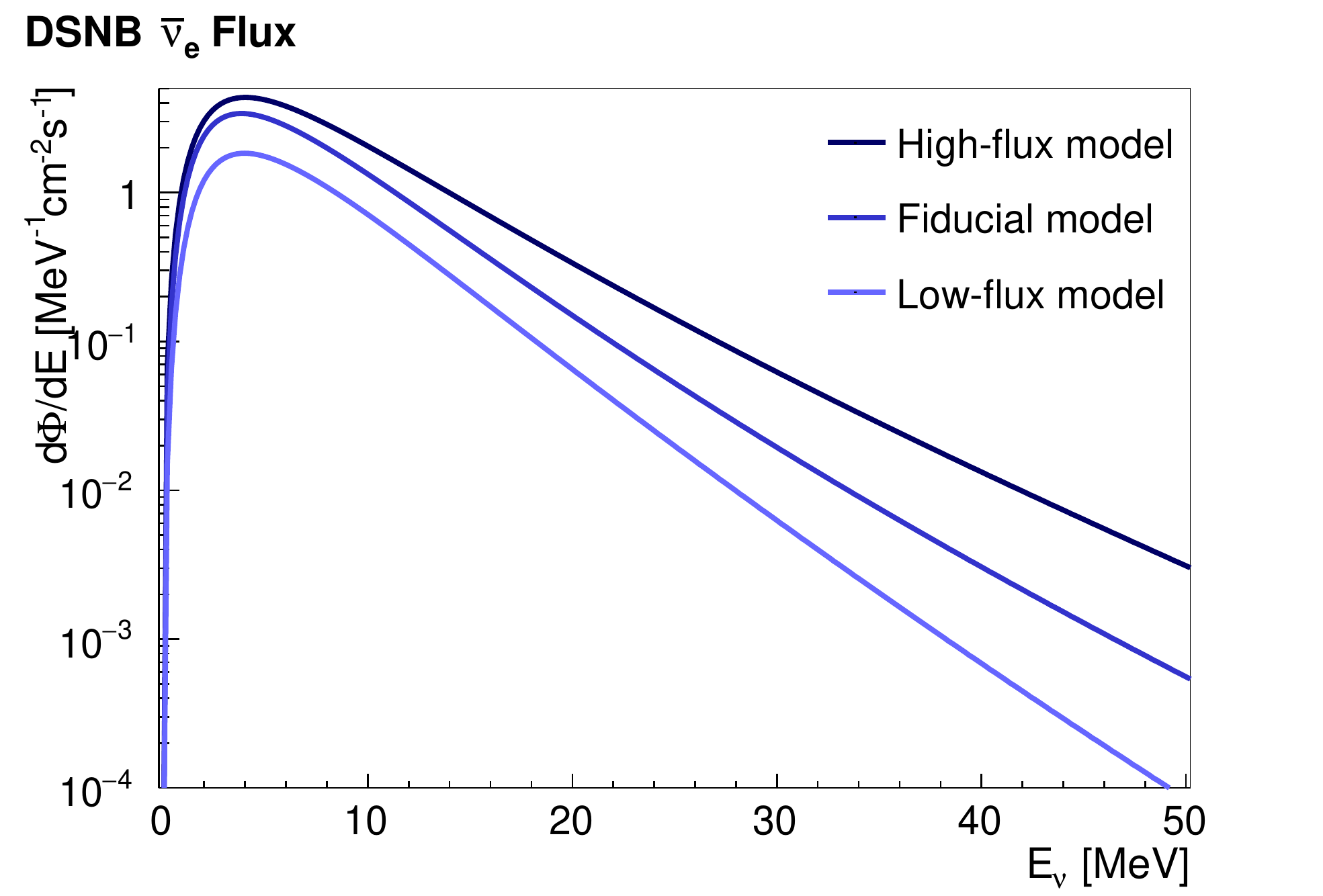}
\caption{\label{fig:dsnbflux}Differential $\bar{\nu}_e$ DSNB flux arriving at Earth with neutrino energy $E_{\nu}$ for the fiducial model (``W18-BH2.7-$\alpha$2.0'' with best-fit SFH from \cite{Mathews:2014}; blue), lying between the low-flux model (``S19.8-BH2.3-$\alpha$2.0'' with $-1\sigma$ SFH from \cite{Mathews:2014}; light blue) and the high-flux model (``W20-BH3.5-$\alpha$2.0'' with best-fit SFH from \cite{Mathews:2014}; dark blue). See Ref. \cite{Kresse:2020} for a more detailed description of the flux models.}
\end{figure}

\textbf{DSNB spectrum.} Given the vast multitude and distance of SNe that contribute to the DSNB, its flux is nearly isotropic.\ The energy spectrum averages over the entire population of stellar core collapses from a wide range of progenitor stars (including failed explosions leading to the formation of BHs) and is substantially red-shifted for sources at far-out distances.\ The detectable signal above \SI{\sim10}{MeV} is thus dominated by relatively close-by core-collapse events up to redshifts of $z \approx 1$ (see, e.g., Ref.~\cite{Ando:2004hc}).\ The DSNB spectrum in a detector is given by
\begin{equation}
\dv{N_{\nu}}{E_{\nu}} = N_p \times \dv{\Phi_{\nu}}{E_{\nu}} \times \sigma_{\nu}(E_{\nu}),
\label{Eq:DSNBSpectrum} \end{equation}
where $\diff\Phi_{\nu}/\diff E_{\nu}$ is the differential DSNB flux, $\sigma_{\nu}(E_{\nu})$ is the energy-dependent cross-section for the IBD reaction \cite{Strumia:2003zx} and $N _p=\num{6.73d32}/(\SI{10}{kt})$ is the number of protons in the target volume. As the Earth is virtually transparent to low-energy neutrinos, detectors see a fully isotropic signal.\ The differential number flux $\diff\Phi_{\nu}/\diff E_{\nu}$ can be computed via a line-of-sight integral of the average SN neutrino number spectrum $\diff N(E'_{\nu})/\diff E'_{\nu}$ (weighted by an initial mass function), multiplied by the evolving core-collapse SN rate $R_\mathrm{SN}(z)$ over the cosmic history \cite{Ando:2004hc,Beacom:2010kk}:
\begin{equation}
\dv{\Phi_{\nu}}{E_{\nu}} =c \int_0^{\infty} \dv{N(E'_{\nu})}{E'_{\nu}} \times \dv{E'_{\nu}}{E_{\nu}} \times R_\mathrm{SN}(z) \times \left|\dv{t}{z}\right| \dd z,
\label{Eq:IntegrationDSNB}
\end{equation}
where $c$ is the speed of light, $E'_{\nu}$ is the neutrino energy at emission, and $E_{\nu}=E'_{\nu}/(1+z)$ denotes the neutrino energy upon detection, corrected for the redshift. The term $|\diff t/\diff z|$ is given by an underlying cosmological model; it accounts for the expansion history of the Universe and relates $z$ to the cosmic time $t$.
\medskip\\ \newpage
\textbf{Model inputs.}
As our DSNB model, we employ the recent flux predictions by \cite{Kresse:2020}.\ They are based on a large set of spherically symmetric SN simulations with the \textsc{Prometheus-HotB} code \cite{Janka:1996,Kifonidis:2003,Scheck:2006,Ertl:2016} over a wide range of progenitor stars with birth masses between $\sim$9 and 120 solar masses (M$_\odot$) and include cases of BH formation in failed explosions. Their modelling approach of using ``calibrated neutrino engines'' follows the works by \cite{Ugliano:2012,Ertl:2016,Sukhbold:2016}. For most parts of our analysis, we take the ``fiducial'' model of \cite{Kresse:2020} (see their Sec.\,4). However, to account for the large uncertainties which are still underlying any theoretical prediction of the DSNB, we additionally consider a ``low-flux'' and a ``high-flux'' model from \cite{Kresse:2020}.\ The flux spectra of the three different models are shown in fig.\,\ref{fig:dsnbflux}. The underlying model parameters are as follows:
\begin{itemize}
    \item \textbf{Fiducial model:} This model with ``Z9.6\,\&\,W18'' neutrino engine includes a 26.9\% fraction of BH-forming, failed SNe; a maximum baryonic neutron star mass of 2.7\,M$_\odot$ was assumed (which corresponds to $\sim2.2$\,M$_\odot$ of gravitating mass); a spectral shape according to \cite{Keil:2003} was used for their time-dependent neutrino signals, which was adjusted to sophisticated SN simulations with detailed microphysics; the core-collapse SN rate scales with the cosmic star-formation history (SFH) for which the parametric description by \cite{Yuksel:2008} was taken together with the best-fit parameters by \cite{Mathews:2014}.
    \item \textbf{Low-flux model:} This model makes for a rather conservative flux prediction and is taken as a lower limit in our study. It is based on SN simulations with the ``Z9.6\,\&\,S19.8'' neutrino engine, leading to a 17.8\% fraction of BH-forming, failed SNe; a maximum baryonic (gravitational) neutron star mass of 2.3\,M$_\odot$ ($\sim2.0$\,M$_\odot$) was used; the same assumptions for the spectral shape as in the fiducial model were taken, whereas the $-1\sigma$ lower-limit parameters by \cite{Mathews:2014} were employed to the cosmic SFH.
    \item \textbf{High-flux model:} This model is chosen such that its integrated flux above $\SI{17.3}{MeV}$ roughly coincides with the current SK upper-flux limit \cite{Bays:2011si}. It employs the neutrino engine ``Z9.6\,\&\,W20'' with a 41.7\% fraction of BH-formation cases; a maximum baryonic (gravitational) neutron star mass of 3.5\,M$_\odot$ ($\sim2.8$\,M$_\odot$) was taken; the same assumptions for the spectral shape and for the SFH as in the fiducial model were used.
\end{itemize}
The reader is referred to Ref. \cite{Kresse:2020} for a more detailed description of the DSNB modelling (also including considerations of neutrino-oscillation effects) and the motivation of the entering parameters and assumptions.\footnote{The flux models can be downloaded from the Garching Core-Collapse Supernova Archive (\url{https://wwwmpa.mpa-garching.mpg.de/ccsnarchive/archive.html}) upon request.}
Using these models, the expected number of DSNB $\bar{\nu}_e$ events in a WbLS detector can be computed according to eq.~(1).\ Assuming the fiducial model to be realised in Nature, we expect 31 events per 100\,kt$\cdot$yrs (for $0 \leqslant E_{\nu} \leqslant \SI{40}{MeV}$).\ The more conservative low-flux model would yield 15/(100\,kt$\cdot$yrs), whereas the high-flux model (roughly corresponding to the current SK limit) would translate to 59/(100\,kt$\cdot$yrs). This wide spread of values illustrates the large uncertainties which are still preventing more precise DSNB predictions at the present day. Nonetheless, it also shows the big potential of future DSNB measurements to independently constrain the (yet wide) parameter space. If not stated otherwise, we refer to the fiducial DSNB model throughout our work but comment on the model dependence of our results in sec.\,\ref{sec:sensitivity}.

% -----Background Sources--------
\section{\label{sec:background}Background Modeling}
\begin{figure}[b]
        \centering
        \includegraphics[width=0.5\textwidth]{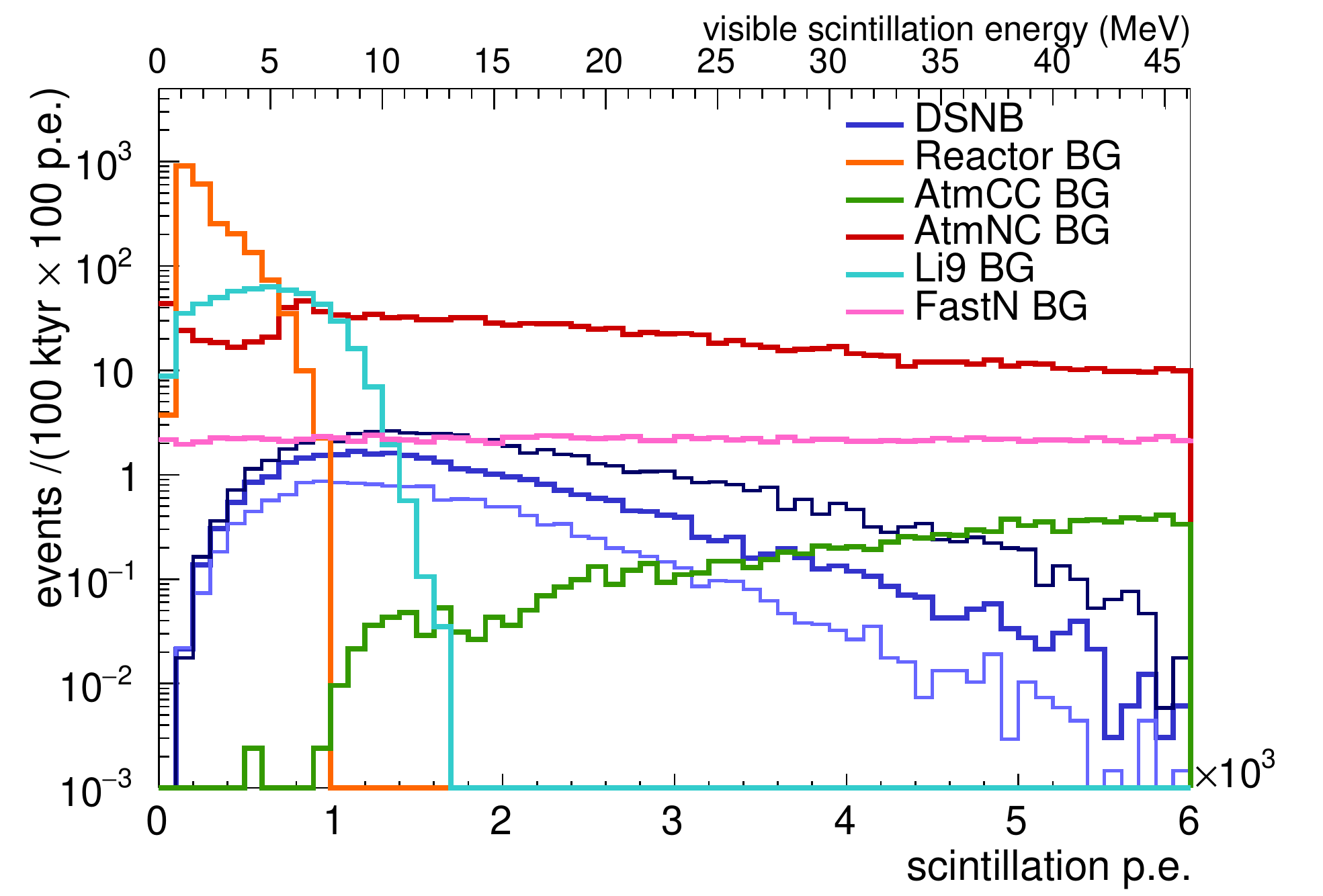}
\caption[]{The visible scintillation energy spectrum expected for the DSNB signal and its ample backgrounds. The presented spectra include reactor neutrinos, cosmogenic \ce{^9Li}, fast neutrons as well as atmospheric neutrino charged-current (CC) and neutral-current (NC) interaction rates. We assume a basic event selection of IBD-like coincidence signals (with only a single accompanying neutron capture). The energy scale is based on the number of scintillation photons detected. The upper axis lists the corresponding visible scintillation energy. Expected rates according to the three DSNB models are indicated in blue (cf.~FIG.\,\ref{fig:dsnbflux}).}
\label{fig:Spectrum_all}
\end{figure}

A variety of backgrounds besets the DSNB signal. They can be divided into three categories: \begin{itemize}
    \item The terrestrial flux of $\bar{\nu}_e$'s from reactors and atmospheric neutrinos causes an irreducible background of real IBD coincidences and reduces the detection window to the range of about 8 to 30\,MeV.
    \item Cosmic muons are penetrating the rock shielding above the detector. In spallation processes, these muons generate $\beta$n-emitting isotopes (e.g.\ \ce{^9Li}) in the target material as well as fast neutrons when passing through the rock surrounding the detector. Both can mimic IBD coincidence signals.
    \item High-energy atmospheric neutrinos undergo neutral-current (NC) interactions in the target medium. In case neutrons are released in these reactions, they can create coincidences with a prompt signal in the visible energy range of the DSNB.
\end{itemize}
In the following, we give a detailed account of the modeling of these background sources. The expected event rates are listed in the first column of tab.\,\ref{tab:dsnb_rates_EventSelection}, while fig.\,\ref{fig:Spectrum_all} displays the corresponding visible energy spectra.

\subsection{Reactor and Low-Energy Atmospheric Neutrinos}
{\bf Reactor neutrinos}.\ $\bar{\nu}_e$'s emitted by nuclear reactors provide a high background flux at energies below 10\,MeV.\ %The $\bar{\nu}_e$ are generated by the $\beta^-$ decay of neutron-rich fission products of \ce{^{235}U}, \ce{^{238}U}, \ce{^{239}Pu} and \ce{^{241}Pu}.\ 
The total reactor neutrino rate (including its uncertainty) and the oscillated energy spectrum are derived from \cite{Barna:2015rza}.\ For the Sanford Underground Research Facility (SURF) in South Dakota, we expect $(2240\pm112)$ events per 100\,kt$\cdot$yrs.
\medskip\\
{\bf Atmospheric neutrinos.} At low energies, the flux of atmospheric $\bar{\nu}_e$'s increases with energy and starts to surpass the DSNB signal at  around \SI{30}{MeV}.\ Since their flux depends on the geographic (geomagnetic) latitude~\cite{Gaisser:1988ar}, we adopt the HKKM atmospheric neutrino fluxes between \SI{100}{MeV} and \SI{d4}{GeV} that have been calculated for the DUNE experiment at the same location \cite{Honda:2015fha}.\ For lower neutrino energies, we extrapolate the FLUKA simulations performed for the location of the Gran Sasso National Laboratories \cite{Battistoni:2005pd} that are at nearly the same geographical latitude ($42.5^{\circ}$\ N vs.\ $44.4^{\circ}$\ N).\ The FLUKA fluxes are scaled to match the atmospheric HKKM spectrum between \SI{100}{MeV} and \SI{200}{MeV}.\ Using the IBD cross-section from \cite{Strumia:2003zx}, the rate of atmospheric neutrino reactions below \SI{100}{MeV} is calculated to ($48 \pm 17)$/(100\,kt$\cdot$yrs).\ The relative error of 35\,\% reflects the uncertainty of low-energy atmospheric flux predictions \cite{Battistoni:2005pd,Guo:2018sno}.
\subsection{Cosmogenic Backgrounds}
%-----LI9-------
 {\bf  In-situ production of \ce{^9Li}.} Cosmic muons create a variety of radioisotopes by spallation on the oxygen (and carbon) nuclei of the WbLS target.\ Of those, only $\beta n$-emitters can mimic the IBD signature and are thus potential contributors to the background.\ The only isotope produced with a relevant cross-section and sufficiently high endpoint energy ($Q=13.6\,$MeV) to reach into the observation window is \ce{^9Li}.\ In $\sim$50\,\% of all cases, \ce{^9Li} decays to an excited state of \ce{^9Be} that de-excites via emission of a neutron
\cite{TableOfIsotopes}.\ The decay scheme of the $\beta^-$-decay of \ce{^9Li} was implemented according to~\cite{Tilley:2002vg,Tilley:2004zz}.\ Since this background scales to first order with the muon flux, a deep location like SURF translates to a substantial reduction in the \ce{^9Li} production rate.\ We estimate the expected background rate by adopting the \ce{^9Li} yield measured for water and the organic component of the WbLS in \ac{SK}
~\cite{Super-Kamiokande:2015xra} and Borexino~\cite{Bellini:2013pxa}, respectively.\ According to \cite{Hagner:2000xb}, we scale this rate to the lower muon flux and higher muon mean energy at SURF, i.e.\ $R_{\ce{^9Li}} \propto \Phi_\mu \cdot \langle E_\mu \rangle^{0.75}$, assuming a muon flux of $\Phi_\mu=\SI{4.2d-9}{cm^{-2}.s^{-1}}$ and mean muon energy of $\left<E_{\mu}\right>=\SI{293}{GeV}$ \cite{Mei:2005gm}.\ The resulting IBD-like background rate is approximately $(530 \pm 106)/(\SI{100}{kt.yr})$, the relative uncertainty of 20\,\% in line with the measured values~\cite{Bellini:2013pxa,Super-Kamiokande:2015xra}.\medskip
 \newline 
%-----FAST N-------
{\bf Fast Neutrons.} High-energy neutrons induced by cosmic muons can mimic the IBD signature.\ The prompt signal is provided by the elastic scattering of the neutron on a free proton in the target material, followed by the thermalization of the neutron and its capture on hydrogen.\ Muons crossing the target volume can be clearly identified, permitting a suppression of the trailing neutron signals by a short veto of the detector following each muon event ($\Delta t\sim1$\,{\textmu s}).\ On the other hand, neutrons induced by muons passing through the rock layer surrounding the detector provide no immediate background tag.\ The reconstructed vertex positions will usually be close to the verge of the detection volume, though.

We estimate the fast neutron production rate based on prior work for the LENA experiment at the Pyh\"asalmi mine \cite{Mollenberg:2014pwa}.\ At this location, rock overburden (4\,km\,w.e.) and mean muon energy of \SI{300}{GeV} are comparable to SURF, so that we scale the rate by a factor 0.5 to take into account the lower muon flux.\ 
Due to the relatively short mean free path of the neutrons ($\lambda_n<\SI{1}{m}$  \cite{Aglietta:1999iw}), the rate of neutrons entering the detector depends on the amount of rock close to the detection volume and is thus proportional to its geometric surface area.\ Since, at this depth, the main component of the cosmic muon flux is close to vertical, only the lateral surfaces of the detector have to be considered.\ For the geometry of Theia100, the resulting fast neutron background rate is $113.2/(\SI{100}{kt.yr})$ for $E_{\rm vis}<\SI{40}{MeV}$.\ Measured neutron yields in Borexino are accurate up to 5\% \cite{Bellini:2013pxa}, but uncertainties on muon flux and neutron propagation will further increase uncertainty on the predicted background rate.\ For Theia25, the resulting fast neutron rate is 60\,\% lower.\ In line with earlier publications from reactor neutrino experiments, we expect an almost flat visible energy spectrum for the neutron recoils (compare e.g.~\cite{An:2013uza}).\ Even if this simplified assumption may not be fully justified, the impact on this study is negligible since the fast neutron background can be reduced very efficiently (sec.\,\ref{sec:bg_reduction}).
\subsection{\label{sec:nc_atmospheric}Neutral-Current Reactions of Atmospheric Neutrinos}
\begin{table}[tb]
\centering
\renewcommand{\arraystretch}{1.3}
\begin{tabular}{lr} \hline
\multicolumn{2}{c}{\ce{\nu_x + ^{16}O -> \nu_x +}}   \\ 
Reaction Channel        & Branching Ratio [\%] \\ \hline\hline
\ce{\text{n} +^{15}O}						        &49.4\\
\ce{\text{n} +\text{p} +^{14}N}				        &20.0	\\					
\ce{\text{n} + 2p + ^{13}C}					        &12.6\\
\ce{\text{n} +	\text{p} + d + ^{12}C}			    &7.7\\
\ce{\text{n} + \text{p} + d + $\alpha$ + ^{8}Be}	&2.1\\
\ce{\text{n} + $\alpha$ + ^3He + ^{8}Be}		    &1.7\\
\ce{\text{n} + 3p + ^{12}B}					        &1.6\\
\ce{\text{n} + \text{p} + $\alpha$ + ^{10}B}		&1.3\\
\ce{\text{n} + 2p +$\alpha$ + ^{9}Be}			    &1.3\\ 
other reaction channels					            &2.3 \\ \hline\hline
\multicolumn{2}{l}{$R_{NC} \simeq$\,12.7/(kt$\cdot$yrs)}	\\ \hline
\end{tabular}		
\caption{Branching ratios of inelastic NC scattering reactions of atmospheric neutrinos on \ce{^{16}O} with one neutron in the final state, including the de-excitation channels, sorted by their prevalence for visible energies below $\sim$\,\SI{46}{MeV}.}
\label{Tab:ReactionsO16}	
\end{table}

The neutral-current (NC) interactions of high-energy atmospheric (anti-)neutrinos of all flavors pose the most serious background to DSNB detection. Events mimicking IBDs originate from reactions with a single neutron in the final state, while the prompt signal can be composed of a multitude of different combinations of nuclear fragments and gamma-rays. First recognized by the KamLAND experiment \cite{Collaboration:2011jza}, it dominates the DSNB signal by more than one order of magnitude. 
\medskip\\
{\bf Expected event rates.} To obtain the event rates in WbLS as a function of final-state particle composition, we use the GENIE Neutrino Monte-Carlo generator (Version 2.12.4) \cite{Andreopoulos:2009rq,Andreopoulos:2015wxa} and feed it with the HKKM atmospheric neutrino spectrum considering a maximum neutrino energy of \SI{10}{GeV}~\cite{Honda:2015fha}.\ The propagation of Cherenkov and scintillation signals created by the final-state particles are obtained using the GEANT4 detector simulation. However, we pre-filter the reactions and include only events that feature at least one neutron in the final state, which is valid for $\sim$\,79\% of all interactions. Since NC reactions on \ce{^{12}C} contribute only 8\,\% of all NC reactions, they are taken into account by increasing the event number resulting from \ce{^16O} interactions by 10\,\%. The overall NC event rate is thus $7.5\cdot10^3$ per 100\,kt$\cdot$yrs below \SI{10}{GeV} of neutrino energy.

Tab.\,\ref{Tab:ReactionsO16} displays the branching ratios for different configurations of nuclear fragments in the final state. Since we are only interested in reactions resulting in a single neutron, the event rate is reduced to a residual of 1270/(100\,kt$\cdot$yrs).

While an expectation for the background event rate can be quoted, it is not straight-forward to determine the corresponding uncertainties on both rate and spectra. The underlying atmospheric neutrino flux is relatively well known up to an uncertainty of less than 10\% in the relevant energy region \cite{Honda:2015fha}. 
%Relatively well known is the uncertainty on the underlying atmospheric neutrino flux of somewhat less than 10\% in the relevant energy region.
However, there is still very little experimental data on the relevant NC cross-sections on oxygen in this energy region \cite{Tanabashi:2018oca}.

In the near future, corresponding data on cross-sections in water can be expected from the ANNIE experiment at Fermilab. ANNIE will provide a high-statistics sample (${\cal O}(10^4)$) of GeV neutrinos from the Booster Neutrino Beam \cite{Back:2017kfo} that will be used to constrain CC and NC cross-sections. A future WbLS phase of ANNIE is currently discussed and would provide even more valuable data in the present context. 
\medskip\\
{\bf Nuclear de-excitations.} When modeling the light output of the prompt events, i.e.\ of the nuclear fragments created in the NC interactions, it is essential to include de-excitation gammas-rays.\ For the present study, we are most interested in NC interactions were a single neutron is knocked out from an \ce{^{16}O} nucleus in the WbLS (to create an IBD-like signature). 

In the simple shell model, the neutron configuration of \ce{^{16}O} is $[(1s_{1/2})^2(1p_{3/2})^4(1p_{1/2})^2]$.\ As the energy of atmospheric neutrinos is large compared to the binding energies of individual nucleons, it is valid to assume that the probability of neutrino interaction does not depend on the nucleon's binding state \cite{Ejiri:1993rh}. When a neutron is knocked out, the interacting nucleus is left with a hole in the corresponding shell. Reordering of the neutron configuration leads to a release of binding energy via particle emission.\ No rearrangement is needed in the 25\,\% of all cases for which the neutron was located in the $1p_{1/2}$ shell. However, with a probability of 50\,\%, the neutron is knocked out from the $1p_{3/2}$ shell.\ In this case, the excited \ce{^{15}O} nucleus undergoes a direct transition to its ground state, causing the emission of a single \SI{6.18}{MeV} gamma-ray (with a branching ratio of $\sim $87\,\%)~\cite{Abe:2019cpx}.\ Finally, knock-outs from the $1s_{1/2}$ shell leave an excited state of \ce{^{15}O} with a potential transition energy that exceeds the separation energy of both protons and neutrons. Thus, de-excitation proceeds mainly via the emission of protons, nucleons and $\alpha$-particles and only sub-dominantly via $\gamma$-rays \cite{Kamyshkov:2002wp}.

Note that there is a chance that the knock-off nucleon will scatter on other nucleons before leaving the nucleus, leading to the emission of additional nucleons. For simplicity, it is assumed that these intra-nuclear scattering reactions only emit nucleons from the $1p_{1/2}$ shell. Moreover, gamma emission can be as well induced in secondary interactions of knock-out protons and neutrons on neighboring \ce{^{16}O} nuclei (sec.\,\ref{sec:bg_reduction}) that may in turn contribute to the overall prompt signal.

Since neither GENIE nor GEANT4 provides details on particles emitted in the de-excitation of atomic nuclei in neutrino interactions, the nuclear reaction program TALYS (Version 1.4) \cite{Koning:2012zqy} was used to describe the particle emission from residual excited nuclei (\ce{^{15}O},\ce{^{15}N}, \ce{^{14}N}, \ce{^{14}C}, \ce{^{13}C}). 

% -----Background Reduction--------
\section{\label{sec:bg_reduction}Background Reduction}
As shown in tab.\,\ref{tab:dsnb_rates_EventSelection}, several of the background rates in a WbLS target significantly exceed the DSNB signal rate.\ Like in water Cherenkov and organic liquid-scintillator detectors, the IBD interactions induced by the reactor and low-energy atmospheric $\bar\nu_e$'s prove an irreducible background and effectively limit DSNB detection to the $\sim$\,[8$-$30]\,MeV range. However, within the observation window, the enhanced event discrimination capabilities of a WbLS detector enable an effective reduction of background rates to a level significantly lower than that of the DSNB signal (sec.\,\ref{sec:sensitivity}). In the following, we discuss the corresponding event selection in detail.
\subsection{\label{sec:basiccuts}Basic Selection Cuts}
\begin{figure}[tb]
        \centering
        \includegraphics[width=0.5\textwidth]{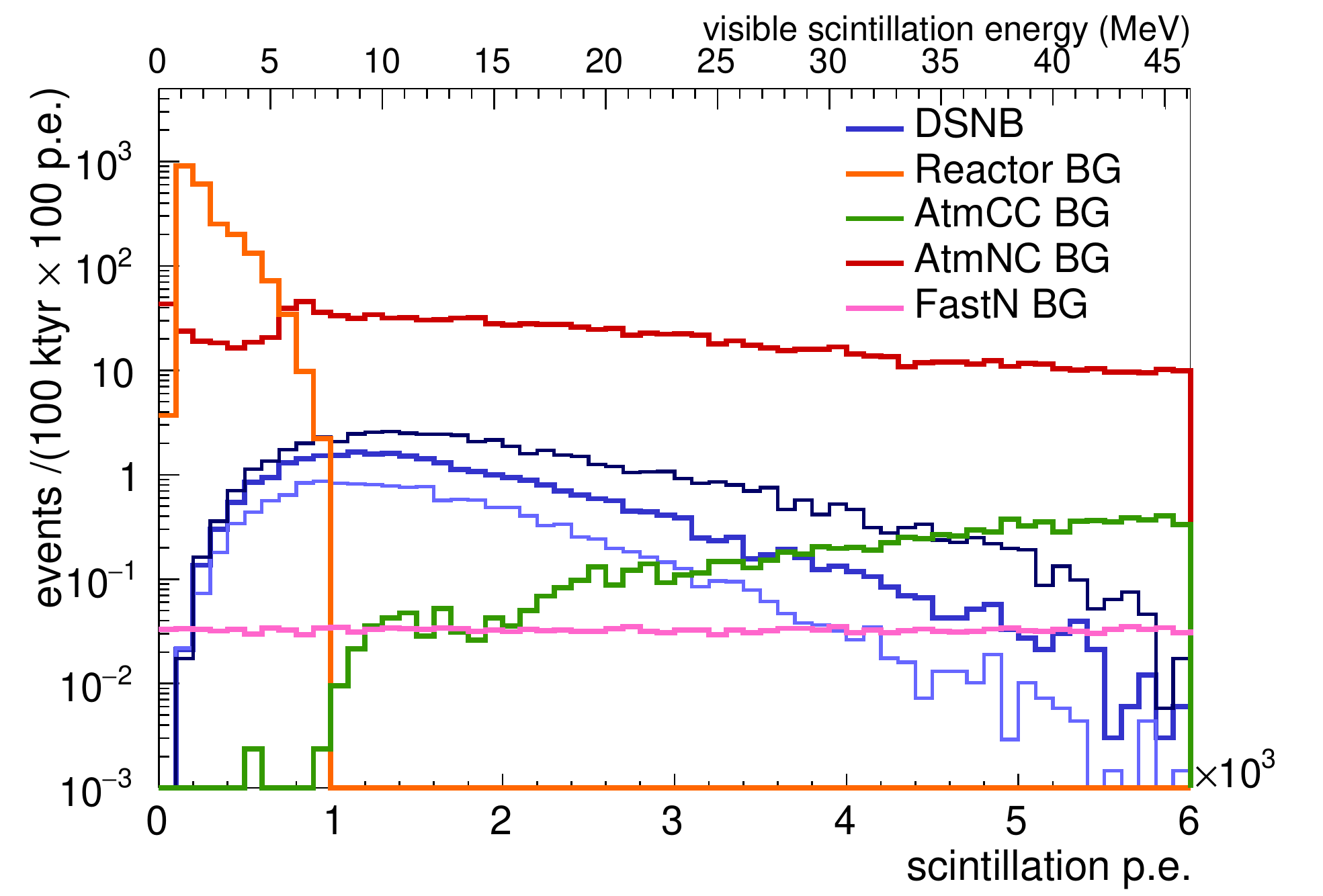}
\caption[]{The visible scintillation energy spectra expected for the DSNB signal and backgrounds after the application of basic discrimination techniques: We apply a \SI{2.5}{m} fiducial volume cut to reduce the background by fast neutrons and a veto of \ce{^9Li} based on the coincidence with the preceding parent muons. Expected rates according to the three DSNB models are indicated in blue (cf.~FIG.\,\ref{fig:dsnbflux}).}
\label{Fig:Spectrum_all_EventSelection}
\end{figure}

{\bf IBD selection.} The analysis of signal and background events starts with the imposition of the selection criteria for the inverse beta decay (IBD). We limit the acceptance of prompt events to the energy range from 8 to 30\,MeV (corresponding to $10^3-4\cdot10^3$\ p.e.\ in Theia100). The lower energy threshold excludes the reactor neutrino background, which is thus neglected in the following. For simplicity, we assume that in this energy region, spatial and time difference cuts applied for selection of the IBD coincidence can be chosen sufficiently wide so that the corresponding efficiency will be close to unity. This assumption is justified since accidental backgrounds will be negligible based on the high energy of the prompt event; the scintillation signal of the 2.2-MeV signal from delayed neutron capture will be well above threshold; and finally, we impose a substantial fiducial volume cut (see below) that will mitigate as well border effects for the IBD selection. It is important to note that the IBD selection only accepts events with exactly one delayed neutron event. As pointed out in sec.\,\ref{sec:nc_atmospheric}, this results in a significant reduction of the NC atmospheric background rate, where multiple neutrons are often generated in the final state.
\medskip\\
{\bf Cosmogenic background veto.} Background from cosmogenic radioisotopes, especially of the $\beta n$-emitter \ce{^9Li}, can be effectively reduced by the formation of a coincidence veto with the preceding parent muon. 
We choose a time veto of \SI{2}{s} following each muon and a cylindrical cut around the muon track of \SI{5}{m} radius. This provides excellent efficiency when compared to the \ce{^9Li} life time of $\tau(\ce{^9Li})=257\,$ms and the expected lateral production profile \cite{Grassi:2015fua}. Given the deep location at SURF, the corresponding loss in exposure is less than 1\,\%. In the following, we regard any possible residuals of \ce{^9Li} as negligible.
\medskip\\
{\bf Fiducial volume cut.} Finally, we impose a fiducial volume cut to reduce surface backgrounds, border effects in detection efficiency and, most crucially, the background imposed by fast neutrons. Given the difference in detector geometries, we choose individual cuts for Theia100, where an outer layer of \SI{2.5}{m} is excluded, while an outer envelope of \SI{1.5}{m} is rejected for Theia25, which reduces the fiducial volume by 20\%.
\medskip\\
The resulting event spectrum of signal and background components after the basic selection cuts is displayed in fig.\,\ref{Fig:Spectrum_all_EventSelection}.
\begin{table}[tb]
\begin{center}
\begin{tabular}{lccc}
\hline
                      &\multicolumn{3}{c}{100\,kt$\cdot$yrs exposure}   \\ \hline
            		& Before Cuts & Muon Veto & Volume Cut  \\
\hline
DSNB signal				    & 30.5	& 29.9	& 29.9\\
Reactor neutrinos		    & 2240	& 2218 	& 2218 \\
Atmospheric CC			    & 9.0	& 8.9	& 8.9\\
\hline
Atmospheric NC			    & 1270	& 1253 	& 1253 \\
$\beta n$-emitters ($^9$Li)	& 529 	& $-$	& $-$ \\
fast neutrons				& 131 (57) 	& 129 (56)  & 2 (4) \\
\hline
\end{tabular}
\end{center}
\caption{Rates of \ac{DSNB} signal and backgrounds at energies below $\sim$\,\SI{46}{MeV} (6000\,p.e.) normalized to a live exposure of 100\,kt$\cdot$yrs.\ The fast neutron rates displayed in the last row assume a \SI{2.5}{m} (\SI{1.5}{m}) fiducial volume cut for Theia100~(Theia25).}
\label{tab:dsnb_rates_EventSelection}
\end{table}
\subsection{\label{sec:ringcounting}Cherenkov Ring Counting}  
\begin{figure}[tb]
        \centering
        \includegraphics[width=0.5\textwidth]{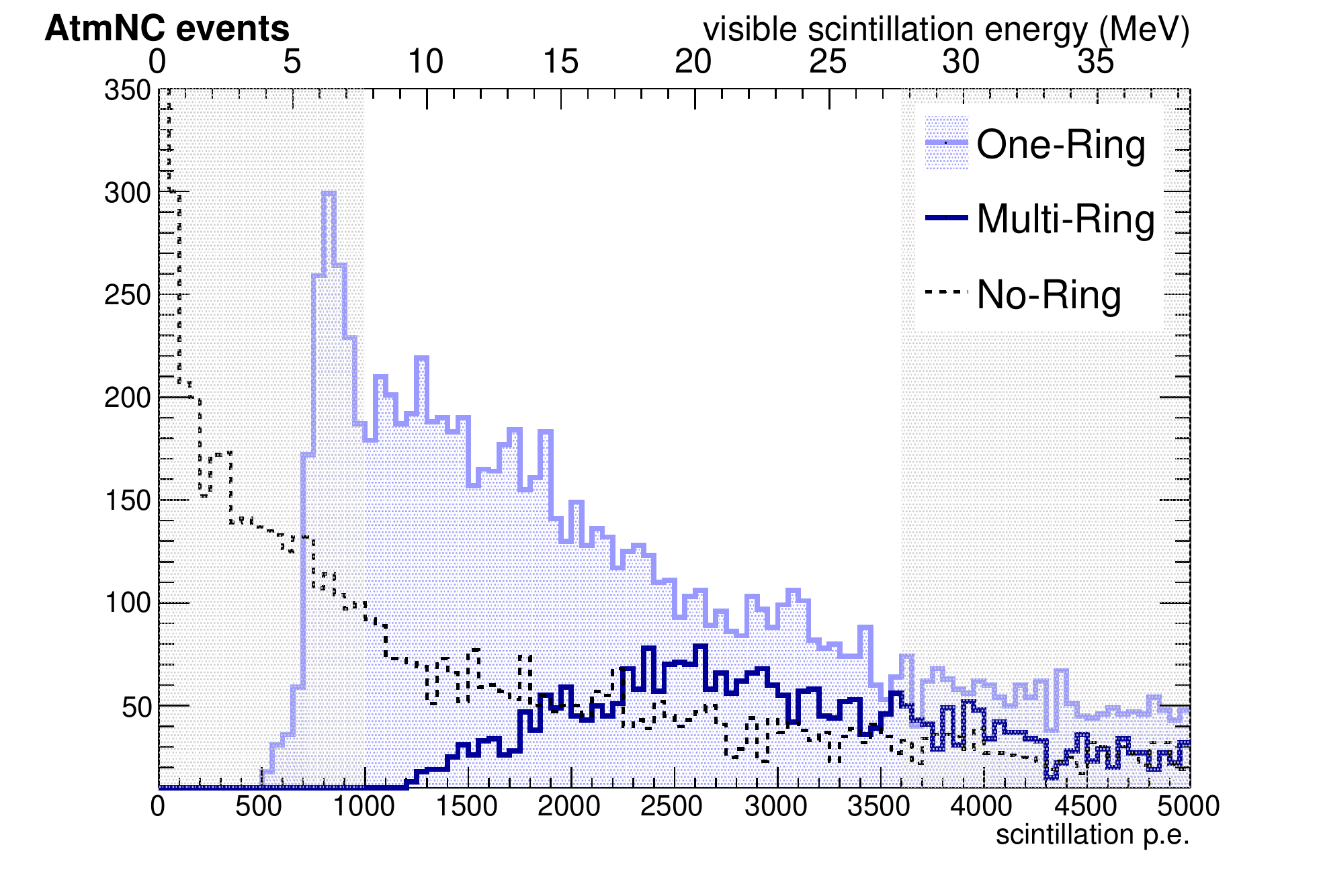}
\caption[]{Number of Cherenkov rings reconstructed for atmospheric NC events as a function of the prompt visible scintillation energy. The frequent occurrence of two or more rings allows for an efficient discrimination against the single-ring DSNB positrons. The grey boxes indicate the limits of the observation window.}
\label{Fig:RingCounting}
\end{figure}

The number of Cherenkov rings created by the prompt event provides a handle to discriminate the DSNB signal against background events that feature multiple particles in the final state. A positron induced by an IBD interaction will create a single Cherenkov ring. In contrast, NC interactions of atmospheric neutrinos often result in several particles above the Cherenkov threshold and thus multiple rings.\ Fig.\,\ref{Fig:RingCounting} displays the number of Cherenkov rings in NC events as a function of the visible scintillation energy. Inside the observation window ($1,000-3,600$\,p.e.), about half of the NC events feature either no or more than a single ring.

To use this approach for event discrimination, the individual Cherenkov rings must be sufficiently bright for reconstruction.\ While a detailed event reconstruction goes beyond the possibilities of this study, we impose here the rather conservative condition that a Cherenkov ring must contain at least 300 p.e.\ to be discernible in the presence of other rings. This condition is immediately met by all signal events, translating to no relevant loss of signal efficiency. Moreover, our analysis code assigns all Cherenkov photons emitted by secondary particles to the initial particles created in the interaction vertex. Thus, secondary particles cannot be discerned as individual Cherenkov emitters and do not enter the ring count. Despite this very conservative assumption,  single-ring selection rejects 57\,\% of atmospheric-neutrino NC interactions as multi-ring background without any relevant loss in signal efficiency.
\subsection{\label{sec:csratio}Cherenkov/Scintillation Ratio}
\begin{figure}[tb]
        \centering
        \includegraphics[width=0.5\textwidth]{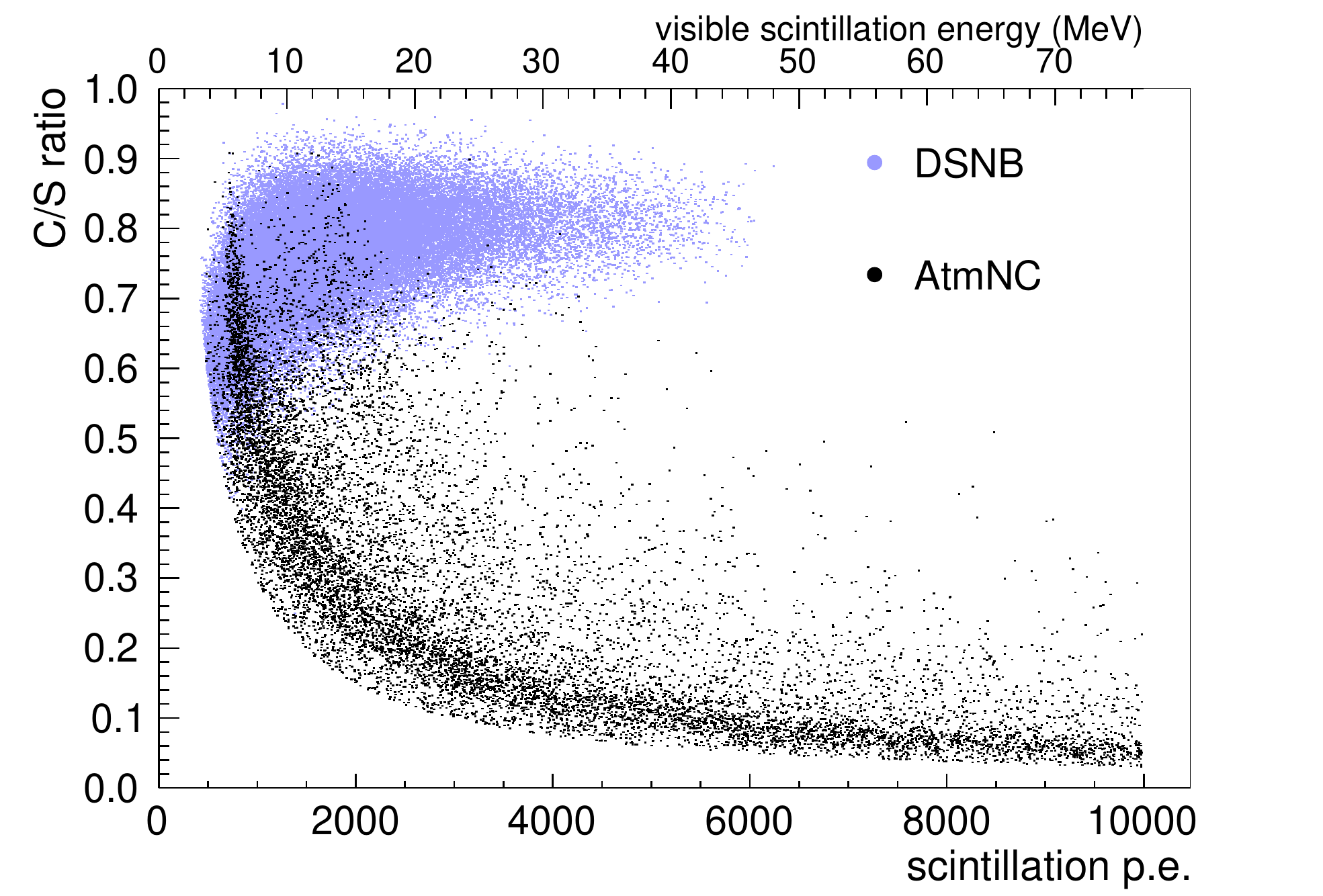}
        \includegraphics[width=0.5\textwidth]{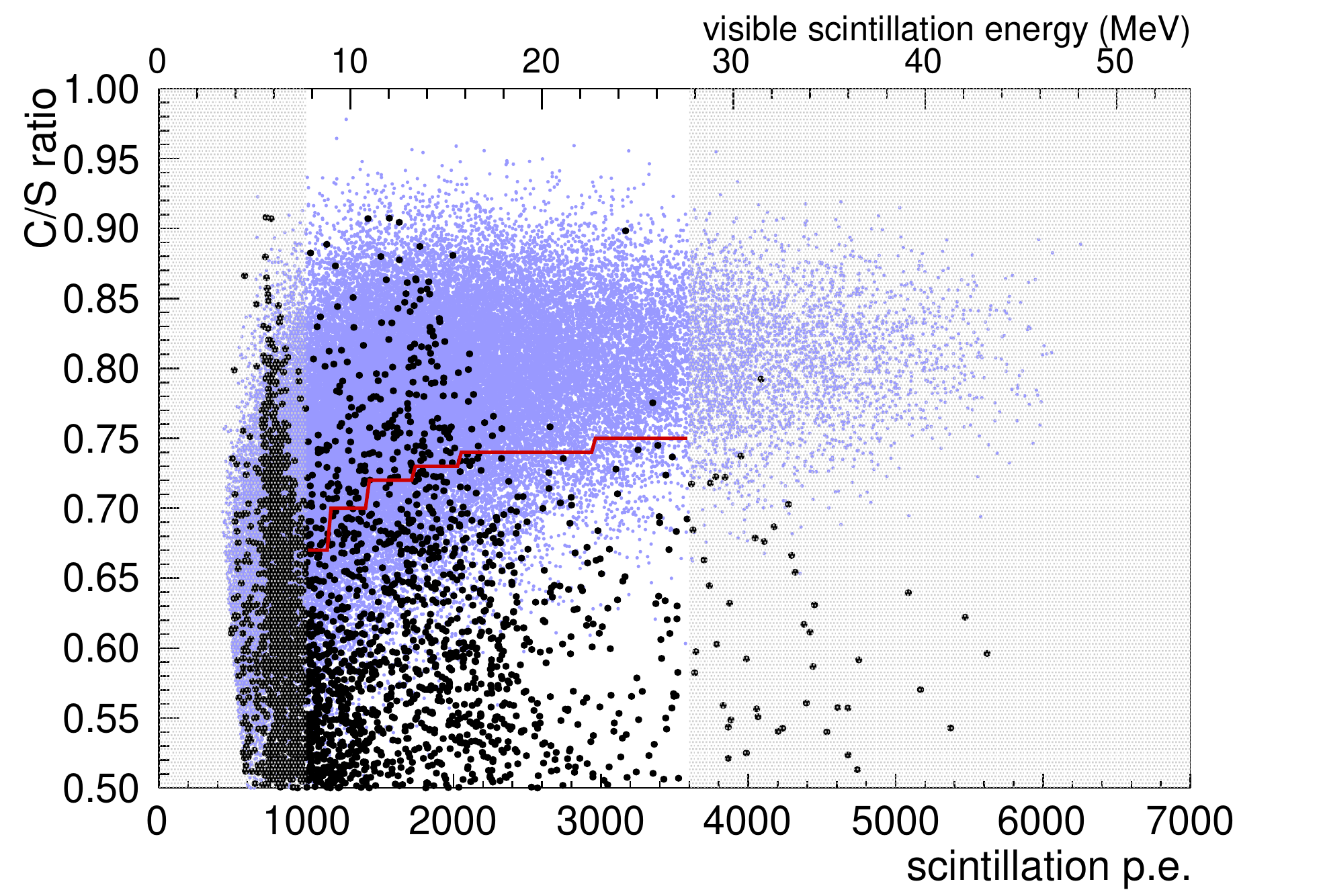}
\caption[]{The Cherenkov-to-scintillation (C/S) ratio offers a powerful tool to discriminate prompt positrons of DSNB events (blue) and hadronic prompt events of atmospheric NC reactions (black).\ Atmospheric NC events lead to a significantly reduced emission of Cherenkov photons.\ The lower plot presents a zoom-in for C/S values greater than 0.5.\ The gray shaded area indicates the limits of the observation window.\ The red line corresponds to the C/S cut threshold reaching 82\% signal efficiency.}
\label{Fig:CSvsScin}
\end{figure}
\begin{figure}[tb]
        \centering
        \includegraphics[width=0.5\textwidth]{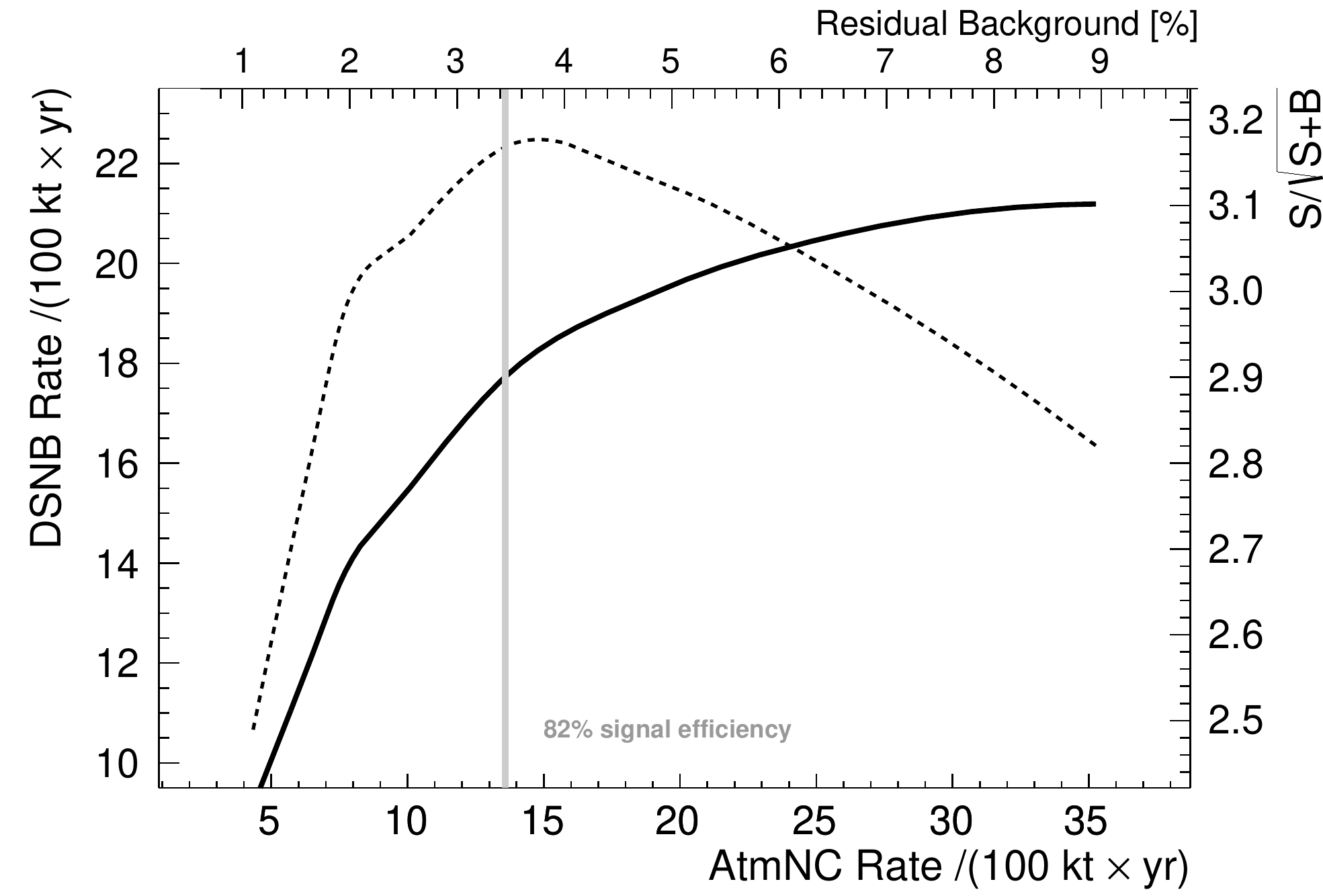}
\caption[]{The optimum choice of a cut on the C/S ratio depends on the optimization of the signal-to-background (S/B) ratio.\ The rate of surviving DSNB events as a function of the residual rate of atmospheric NC events is indicated by the solid line.\ While not shown, the corresponding C/S cut threshold is increasingly relaxed from left to right.\ The dashed line indicates the corresponding significance of the signal over background $S/\sqrt{S+B}$ (scale on the right y-axis).\ The maximum of the curve (82\% signal efficiency at 3.5\,\% residual background, indicated by the grey line) is chosen for the further analysis.}
\label{Fig:CSvsScin2}
\end{figure}

The discrimination technique unique to WbLS detectors is the evaluation of the ratio of reconstructed Cherenkov to scintillation photons, named C/S ratio in the following.\ This parameter depends crucially on the particle type: While the relativistic prompt positrons of IBD events feature a high C/S ratio, the final states of atmospheric NC reactions are mostly hadronic and emit no or comparatively little Cherenkov light.\ This holds even more for the proton recoils induced by fast neutrons in the WbLS target.\ The resulting discrimination power for IBDs and atmospheric NC interactions is demonstrated in fig.\,\ref{Fig:CSvsScin}.\ It displays the C/S ratio as a function of visible scintillation energy.\ In the range above \SI{8}{MeV} ($10^3$\,p.e.), there is a clear separation of the event distributions.\ The residual NC background contamination arises from NC reactions with one or more $\gamma$-rays in the final state (sec.\,\ref{sec:nc_atmospheric}).\ The prevalence of $\gamma$-rays of $\sim$\,\SI{6}{MeV} creates an easy-to-spot curved band in the distribution of atmospheric NC events.\ The lower panel of fig.\,\ref{Fig:CSvsScin} represents a zoom-in of the signal region for C/S values greater than 0.5.\ In this energy range, NC events exhibiting only a single \SI{6}{MeV} gamma-ray feature a lower C/S ratio and are thus automatically excluded.\ However, a second population of events is discernible that likely features two gammas.\ Moreover, there is a couple of events at higher visible scintillation energies that are caused by secondary interactions of high-energy neutrons with other nuclei.\ Since both event populations are expected to create more than a single Cherenkov ring, it is likely that these events $-$ though not discernible by their C/S ratio $-$ could be identified as background by Cherenkov ring counting (see above).\ However, since the current study lacks a sufficiently detailed modeling and reconstruction of these events, we conservatively assume that they cannot be discriminated.

In order to utilize the C/S ratio in the analysis, we define an energy-dependent lower threshold for signal selection.\ It is exemplified by the red line inserted in fig.\,\ref{Fig:CSvsScin}.\ The exact threshold values imposed depend on the focus of the DSNB analysis.\ Here, we optimize for the detection potential of the DSNB, i.e.\ we apply an energy-dependent selection cut that maximizes the significance of the DSNB signal $S$ over the NC background $B$, represented by the signal significance $S/\sqrt{S+B}$ and visualized in fig.\,\ref{Fig:CSvsScin2}.\ For the shown configuration, we reach an optimum in DSNB signal acceptance of 82\,\%, leaving a residual of only 3.5\,\% for atmospheric NC background events and  $\sim 1\%$ for fast neutron events (not shown).

Given the importance of this discrimination technique for the DSNB detection, we also investigated its dependence on the light collection.\ Since Theia25 plans for an initial coverage of 25\,\%, the number of photo electrons collected would be reduced by a factor three compared to 75\,\% coverage of Theia100.\ While the larger uncertainty in photon statistics translates to a slight weakening of the discrimination power, the effective S/B values are only mildly affected: for the optimum threshold, the signal acceptance is 78\,\% while a background residual of 3.7\,\% is permitted.
\subsection{\label{sec:delayeddecays}Delayed Decays}
As displayed in tab.\,\ref{Tab:ReactionsO16}, a significant fraction of the NC reactions of atmospheric neutrinos on \ce{^{16}O} (with a single neutron in the final state) leaves behind a radioactive isotope.\ In principle, its decay can be used to reject the original IBD-like interactions by means of a delayed coincidence tag. In water, the dominant isotope created is \ce{^{15}O} with a branching ratio of $\sim$49\%. \ce{^{15}O} undergoes a $\beta^+$-decay with an endpoint of \SI{2.8}{MeV} and a lifetime of \SI{2.9}{min}
\cite{AjzenbergSelove:1991zz}. For both detector configurations of Theia, this will produce a sizable signal visible at high efficiency. 

Given the relatively long lifetime of \ce{^{15}O} and low decay energy, a refined selection condition must be applied for the delayed coincidence to prevent a high rate of accidental coincidences of DSNB neutrino events with radioactive decays intrinsic to the scintillator.\ In accordance with \cite{Askins:2020aa}, we assume a contamination of the water with elements of the \ce{^{238}U}/\ce{^{232}Th} chains on the level of $10^{-15}$\,g/g. To reduce the rate of accidental coincidences, we impose a maximum delay time of \SI{10}{min} (about 3 life times of \ce{^{15}O}) and a spatial distance of 1\,m between the vertices. This is sufficient to reduce the probability of accidental coincidences of signal events to 1\,\%, while the veto efficiency for delayed-decay NC interactions is still at 95\,\%.

Beyond \ce{^{15}O} and the low-yield isotope (in water) \ce{^{12}B}, none of the other isotopes created permits a similar vetoing technique. While \ce{^14N}, \ce{^13C}, \ce{^12C}, \ce{^10B}, and \ce{^9Be} are stable, \ce{^8Be} almost immediately decays into two alphas.

Arguably, WbLS can be regarded as the only technique that will provide reasonably high efficiency for the delayed-decay veto. For pure Cherenkov detectors, tagging efficiency for \ce{^{15}O} will be very low. In organic scintillators, the main product of NC interactions on \ce{^12C} is \ce{^11C}. While these decays are above the threshold, the lifetime is a long $\sim$\,\SI{30}{min}, making a veto based on the delayed decay as proposed in \cite{PhD_Moellenberg} quite challenging. 

% -----Sensitivity--------
\section{\label{sec:sensitivity}Detection Potential}
\begin{figure}[tb]
        \centering
        \includegraphics[width=0.5\textwidth]{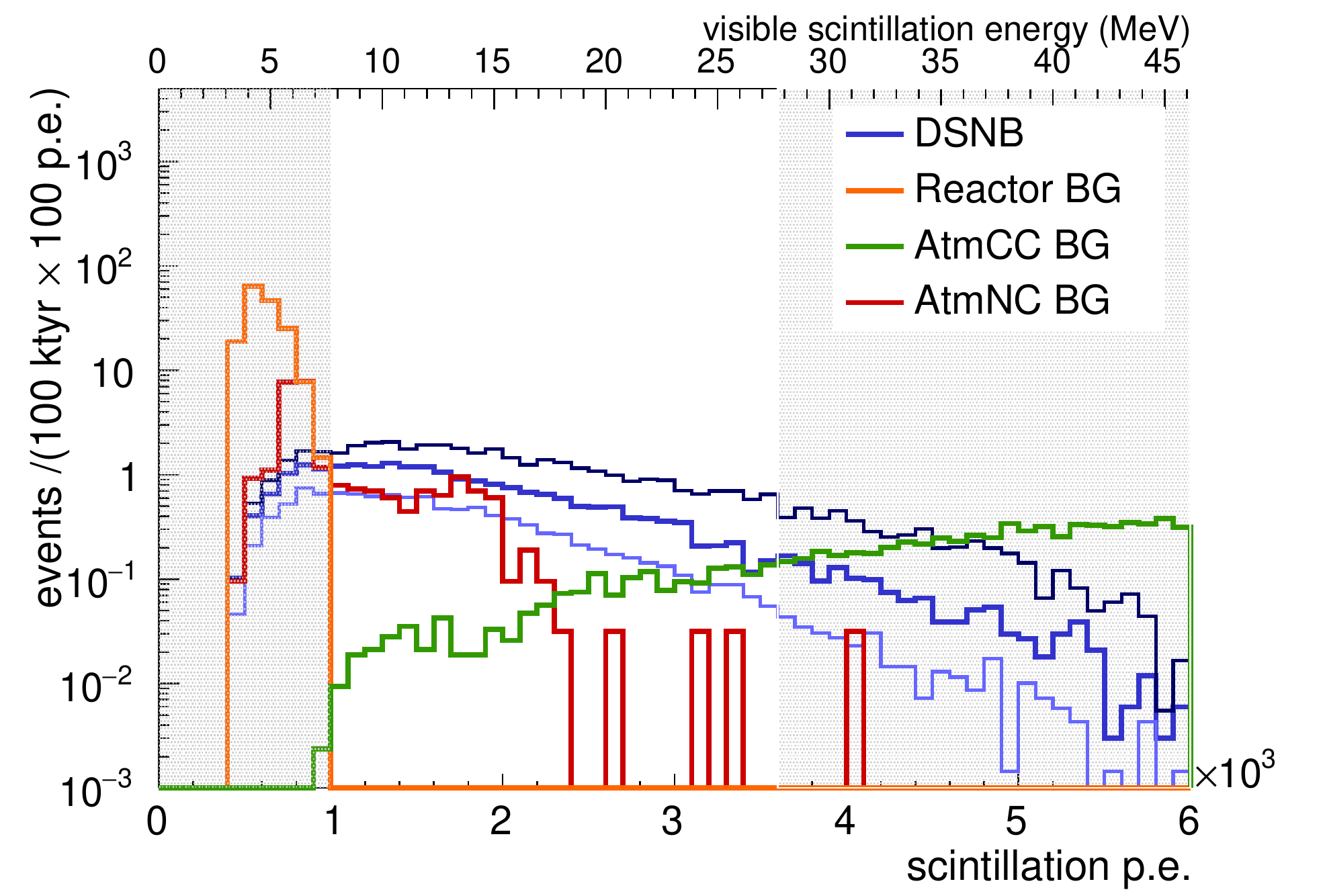}
\caption[]{The visible scintillation energy spectrum expected for DSNB signal and backgrounds after all selection cuts.\ The background components include IBDs from reactor and atmospheric neutrinos as well as a residual of IBD-like NC interactions of atmospheric neutrinos.\ The signal dominates w.r.t.~to the backgrounds over the entire observation window (white region).\ Expected rates according to the three DSNB models are indicated in blue (cf.~FIG.\,\ref{fig:dsnbflux}).}
\label{Fig:Spectrum_all}
\end{figure}

\begin{figure}[t]
        \centering
        \includegraphics[width=0.5\textwidth]{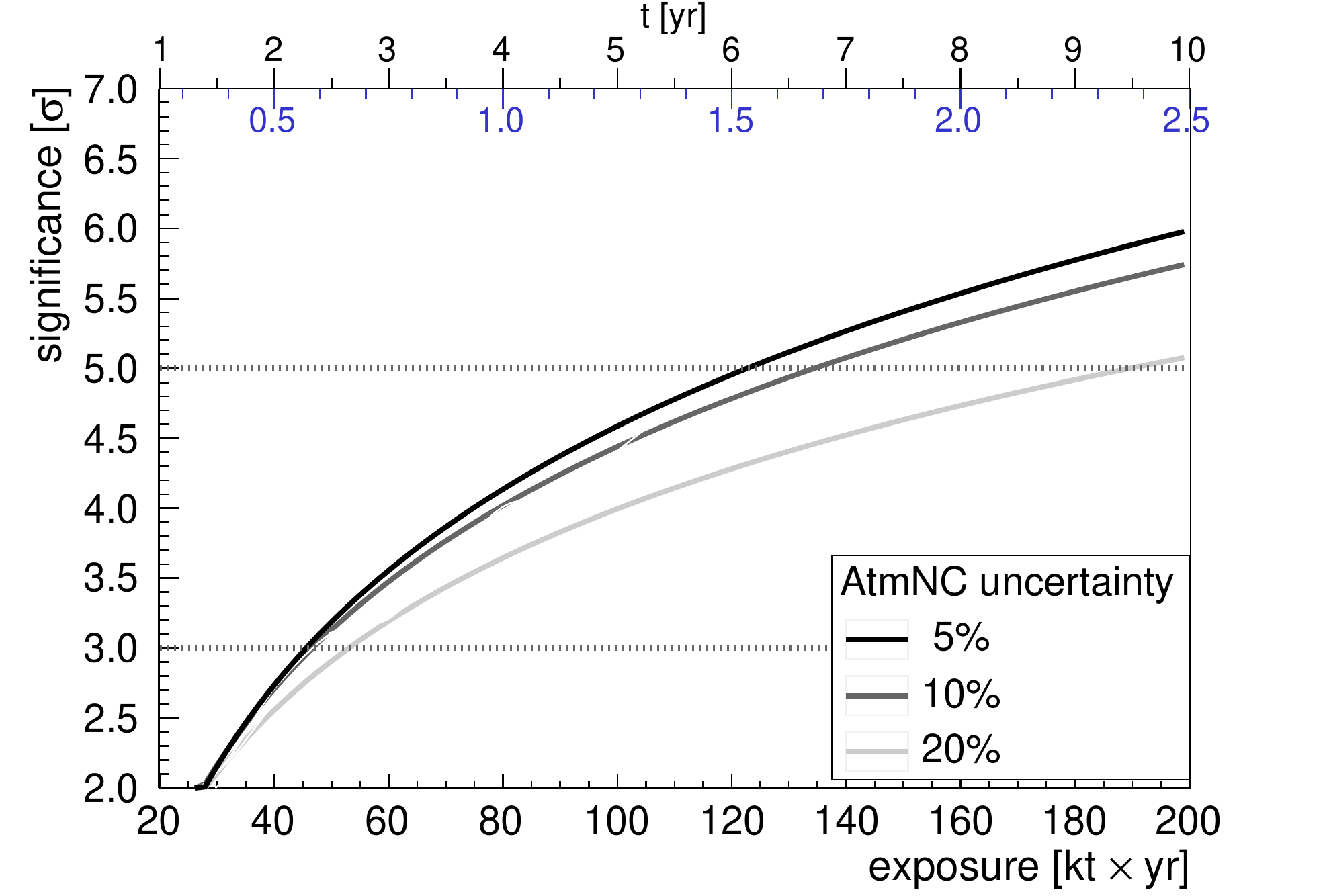}
\caption[]{Significance of \ac{DSNB} detection as a function of exposure.\ The curves correspond to a variation of the relative uncertainty in the atmospheric NC background rates from 5 to 20\%.\ The upper black (blue) horizontal axis scale indicates the operation time of Theia25 (Theia100).}
\label{Fig:Sign1}
\end{figure}

\begin{table*}
\begin{ruledtabular}
\begin{center}
\begin{tabular}{lcccc}
\hline
  &\multicolumn{3}{c}{100\,kt$\cdot$yrs exposure} \\ \hline
Spectral component		& basic cuts & single-ring & C/S cut & delayed decays \\
\hline
DSNB signal				& 21.7	& 21.7	& 17.7 (17.4)	& 17.5 (17.2)	\\ Atmospheric CC			& 2.0		& 2.0		& 1.7 (1.6)		& 1.7 (1.6)	 \\
Atmospheric NC			& 682	& 394 	& 13.6 (14.6)	& 7.4 (7.9)\\
fast neutrons			& 0.8 	& 0.8 & $-$	& $-$\\
\hline
Signal efficiency       & 1     & 1     & 0.82 (0.81)    & 0.81 (0.80) \\
Background residual     & 1     & 0.58  & 0.022 (0.024) & 0.013 (0.014) \\  
Signal-to-background	& 0.03	& 0.05	& 1.2 (1.1)		& 1.9 (1.8) \\
Signal significance     & 0.8   & 1.1   & 3.1 (3.0)     & 3.4 (3.3) \\ 
\hline
\end{tabular}
\end{center}
\caption[Rates of \ac{DSNB} signal and backgrounds]{Integral rates of \ac{DSNB} signal and backgrounds within the observation window (8$-$30\,MeV) for a live exposure of 100\,kt$\cdot$yrs in Theia100 (Theia25).\ The first column represents the rates applying the basic selection cuts described in sec.\,\ref{sec:basiccuts}, where reactor and cosmogenic backgrounds have already been reduced to a negligible level.\ The following columns correspond to the selection of single-rings, application of \ac{C/S} ratio cut, and a delayed-decay veto (secs.\,\ref{sec:ringcounting}-\ref{sec:delayeddecays}).\ Furthermore, the fraction of DSNB event and the residual background retained in the observation window, signal-to-background (S:B) ratio as well as the signal significance over background $S/\sqrt{S+B}$, is given.}
\label{tab:dsnb_rates}
\end{ruledtabular}
\end{table*} 
The excellent background discrimination capabilities of WbLS mean that a detector based on this technique can hope for fast acquisition of statistics on the DSNB signal at comparatively low background levels.\ Tab.\,\ref{tab:dsnb_rates} illustrates the impact of the sequence of event selection cuts that have been introduced in sec.\,\ref{sec:bg_reduction}.\ A great reduction of the overall background levels is evident.\ This is especially true for the rejection of the background by atmospheric-neutrino NC interactions that is reduced by almost two orders of magnitude.\ At the same time, the DSNB signal acceptance is only mildly affected.\ 
\medskip\\
{\bf Signal and background rates.} Overall, $\sim$$(9\pm2)$ background events per 100\,kt$\cdot$yrs remain in the observation window, while the expected signal rate is $\sim$$17_{-8}^{+34}$ events per 100\,kt$\cdot$yrs.\ The range in the signal corresponds to the uncertainty in the DSNB flux prediction.\ The corresponding visible-energy spectrum of signal and background events is shown in fig.\,\ref{Fig:Spectrum_all}.\ The exact signal and background numbers depend substantially on the C/S ratio cut imposed.\ When comparing to the data sample left after basic event selection, we find a final signal efficiency of $>$80\,\% and a background residual of 1.3\,\%.\ The direct signal-to-background (S:B) ratio is 1.9\,(1.8) for Theia100\,(Theia25).\ As described in sec.\,\ref{sec:csratio}, we optimize the signal significance over the background instead, which amounts to \mbox{$S/\sqrt{S+B}\approx3.4$} for an exposure of 100\,kt$\cdot$yrs.\ 
\medskip\\
{\bf Sensitivity of detection.} When calculating the expected significance for the discovery of the DSNB signal, we regard only the total event numbers of signal and background detected in the observation window.\ For better comparability of the various experiments, no spectral information is used.\ The measured DSNB rate is determined by statistical subtraction of the background rates from the overall number of detected events.\ The corresponding confidence interval is calculated according to \cite{Rolke:2004mj}.\ Obviously, this method requires to impose priors on the relevant background rates, mostly based on  control windows:
\begin{itemize}
    \item {\it True IBDs:} The (negligible) contribution of reactor neutrinos can be constrained in the energy range below the observation window, IBDs by atmospheric neutrinos in the energy region above.\ For the latter, we assume a scaling uncertainty of 20\,\%.
    \item {\it $\beta n$ emitters:} The efficiency of the spatial cut surrounding muon tracks can be calculated based on the acquired lateral distribution.\ The contribution is, in any case, negligible.
    \item {\it Fast neutrons} can be constrained based on the expected exponential radial profile and the rate measured in the WbLS target outside the fiducial volume.
     \item {\it Atmospheric NC events:} Due to the large reduction factor for this background, it is hard to estimate from first principles how well the relative uncertainty on the residual rate can be estimated.\ However, both the event rates before and after the application of background rejection cuts can be constrained by extrapolation from the energy window above $\sim$30\ MeV.\   
\end{itemize} 
While it is most likely that the uncertainty on the NC background rate will remain the dominant source of systematic uncertainty, it is hard to constrain its value based on present knowledge.\ Therefore, we decided to leave this quantity as an open parameter in the sensitivity studies, varying the relative uncertainty in a range from 5 to 20\,\% of the predicted NC rate value.\ Fig.\,\ref{Fig:Sign1} displays the DSNB detection significance as a function of the acquired exposure, varying the systematic uncertainty on the atmospheric NC background rates.\ Even under quite unfavorable conditions, a $5\sigma$-discovery can be achieved based on a live exposure of 190\,kt$\cdot$yrs.\ This translates to $\sim$2.4\,yrs (9.5\,yrs) of measuring time for Theia100 (Theia25).
\begin{figure}[tb]
        \centering
        \includegraphics[width=0.5\textwidth]{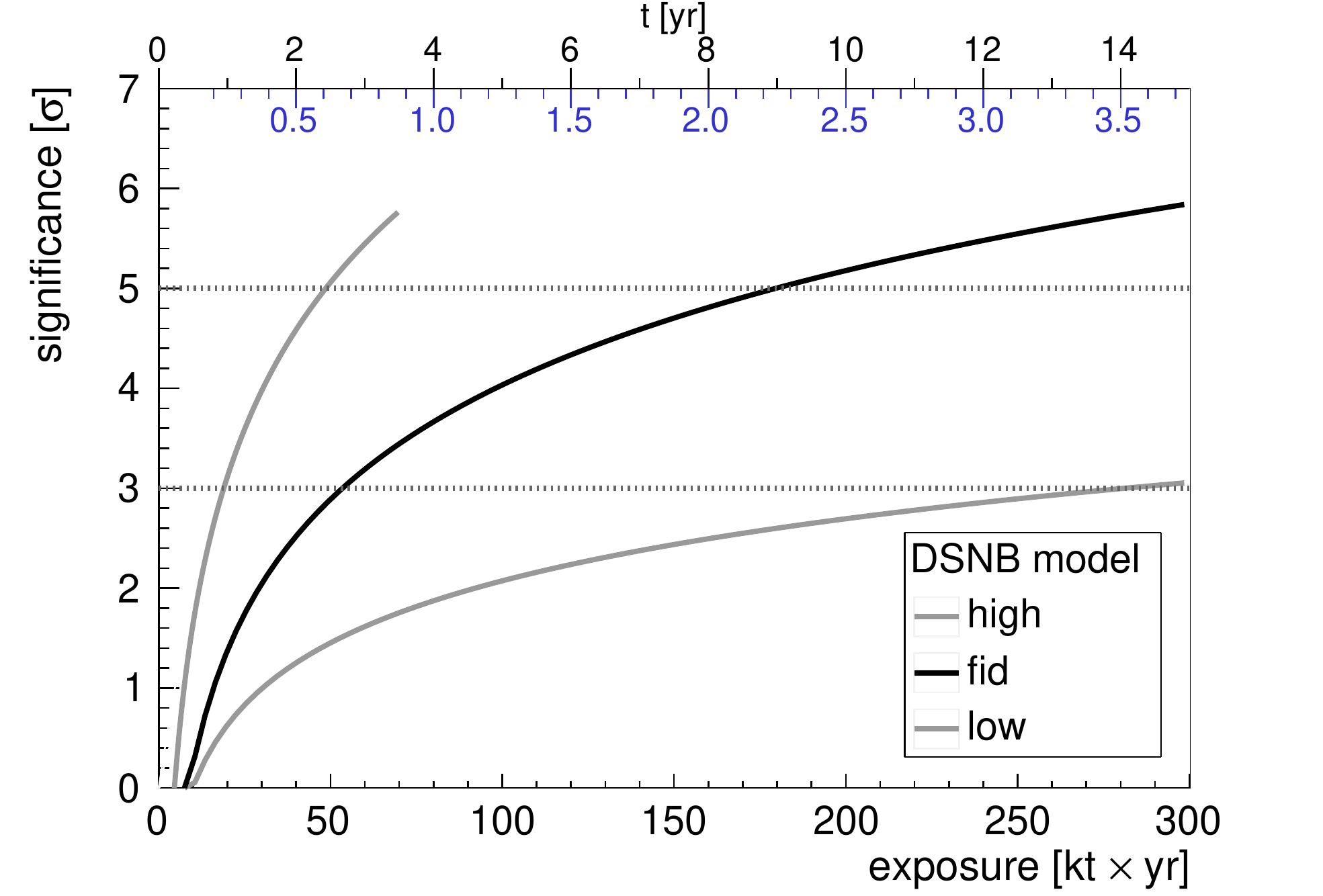}
\caption[]{Significance of \ac{DSNB} detection as a function of exposure for the different DSNB flux models considered in our work.\ The black line corresponds to the fiducial model, whereas the high-flux and low-flux models are given in gray.\ The maximum exposure required for a $3\sigma$-detection (in case of the low-flux model) is $\sim$\,280\,kt$\cdot$yrs.\ The upper black (blue) horizontal axis scale indicates the operation time of Theia25 (Theia100).}
\label{Fig:Sign2}
\end{figure}
\newpage
{\bf Model dependence.} The measuring time required for a positive detection of the DSNB depends strongly on the underlying flux model realized in Nature.\ This dependence is depicted in fig.\,\ref{Fig:Sign2}, where the uncertainty of the predicted atmospheric NC rate is fixed to 20\%.\ While a moderate exposure of only 50\,kt$\cdot$yrs suffices for a discovery in case of a high-flux model (see sec.\,\ref{sec:dsnb}), 280\,kt$\cdot$yrs would be required for a detection at $3\sigma$ level if the DSNB is best described by a low-flux model.\ This is still well within the proposed operation times of both Theia100 and Theia25.
%Especially, the low energy threshold of \SI{8}{MeV} can be used to probe cosmic noon neutrinos (between $1.5\leq z \leq 3$) to pin down the star formation rate \cite{Riya:2020wpw}.
%n a recent work with my student, we have shown that low threshold DSNB
%searches can be used to probe cosmic noon neutrinos (from z ~ 2).\ These
%neutrinos are a probe of the period of peak star formation of the universe
%and can help pin down the true star formation rate, which has a
%discrepancy in current measurements.\ Lowering the threshold to 8 MeV
%should allow a much greater sensitivity to such neutrinos.

%---- others ------
\section{\label{sec:others}Comparison with Other Techniques}
\begin{table*}
\renewcommand{\arraystretch}{1.3}
\begin{ruledtabular}
\begin{tabular}{llrcccrrrc}

Technology	&Experiment			&FM 			&Start 		&Energy Window 	&Signal		&Signal	&BG & \multirow{2}{*}{$\frac{\rm S}{\sqrt{\rm S+B}}$}& t[yr] \\ 
			&					& [kt]				&			&[MeV] 	&Efficiency	&\multicolumn{2}{c}{[/(100\,\si{kt.yr})]} & &for $3(5)\sigma$\\ \hline \hline
	\multirow{2}{*}{WbLS}	&Theia25	&20 	&2030	&\multirow{2}{*}{8$-$30} 	&\multirow{2}{*}{0.8} &17.2 	&9.5    & 3.3 & 4.1 (11.4) \\ 
					        &Theia100   &80		&2035	&	& &17.5	&9.1 & 3.4 & 1.0 (2.7)\\ \hline
LS                          &JUNO       &17.0	&2021	&10.2$-$29.2	&0.5 &9.6	&4.2 &2.6 &7.9 (22.0) \\	\hline				
\multirow{3}{*}{WCD}	 	&SK-Gd	    &22.5	&2021	&10$-$30 	    &0.7 &12.9	&14.0 & 2.5 &6.5 (18.0)\\
					        &HK		    &187	&2027	&20$-$30	    &0.9 &4.0	&39.3 & 0.6 &13.0 (36.2)\\
					        &HK-Gd	    &187	&2033	&10$-$30 	    &0.67 &12.4	&14.0 & 2.4 &0.8 (2.3)\\ \hline

LAr 					&DUNE		& 20+20	&2026 	&16$-$40	&&11.4	&6.0 &2.7 &3.0 (8.4)\\
\end{tabular}
\end{ruledtabular}
\caption{\label{Tab:Comp}Comparison of neutrino observatories aiming at the detection of the {DSNB} signal. The first columns list experimental technique, abbreviation of the experiment ({\it see text}), fiducial mass (FM), projected start of data taking, and observation energy window. For better comparability, the expected signal rate was recalculated for the DSNB fiducial model assumed throughout the paper, using information from  \cite{Talk_Vagins,Talk_Takatomi,Abe:2018uyc,Abi:2018dnh}. Background (BG) rates are scaled from the same sources. The given rates for DUNE are based on \cite{Cocco:2004ac}.\ The corresponding signal significance over background and the required measuring time for $3 (5)\sigma$ is calculated in the last columns. For comparison, we show the expected performance of Theia25 and Theia100. The WbLS technique offers the best signal acceptance and highest signal significance.}
\end{table*}
\begin{figure*}[tb]
        \centering
        \includegraphics[width=0.49\textwidth]{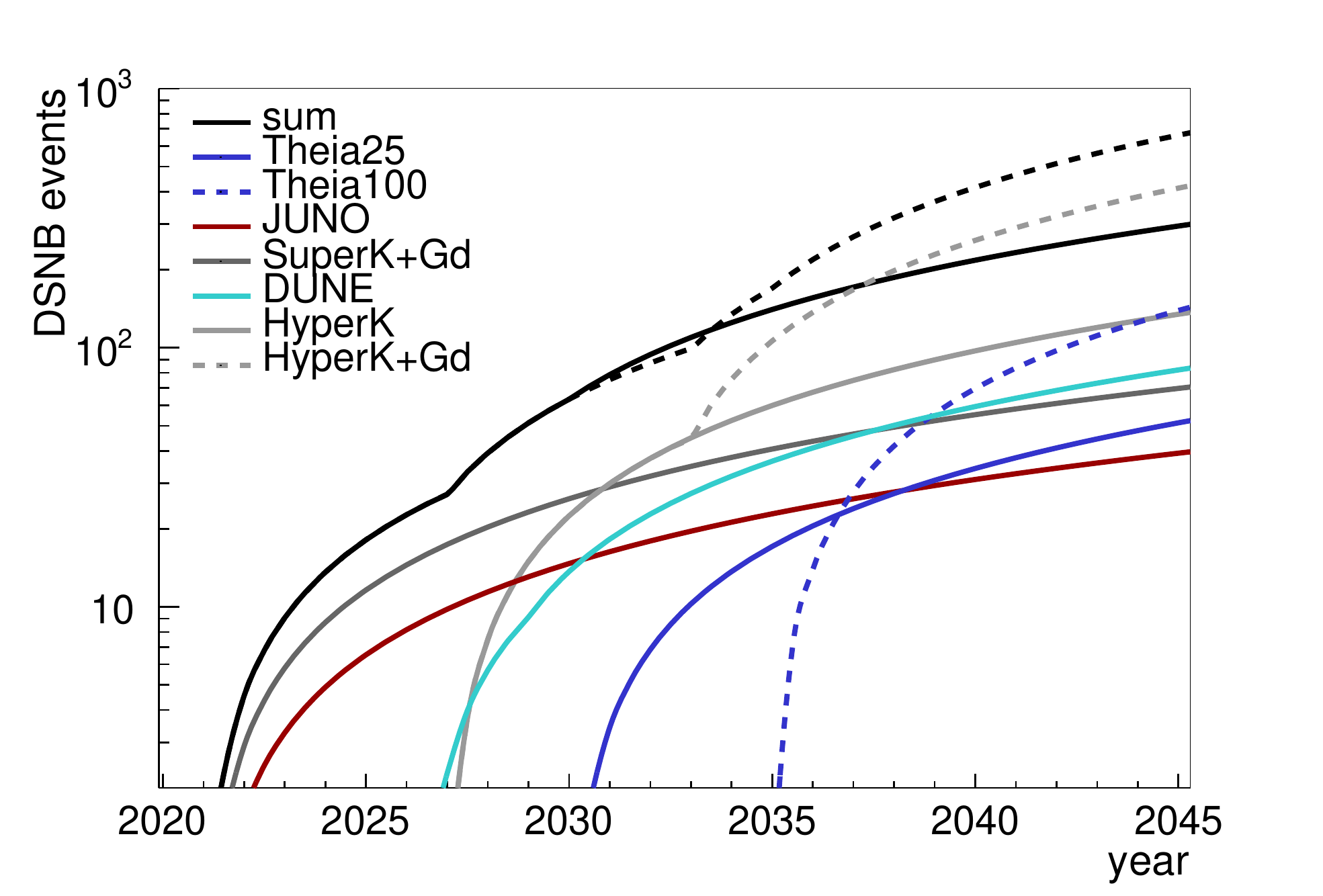}
        \includegraphics[width=0.49\textwidth]{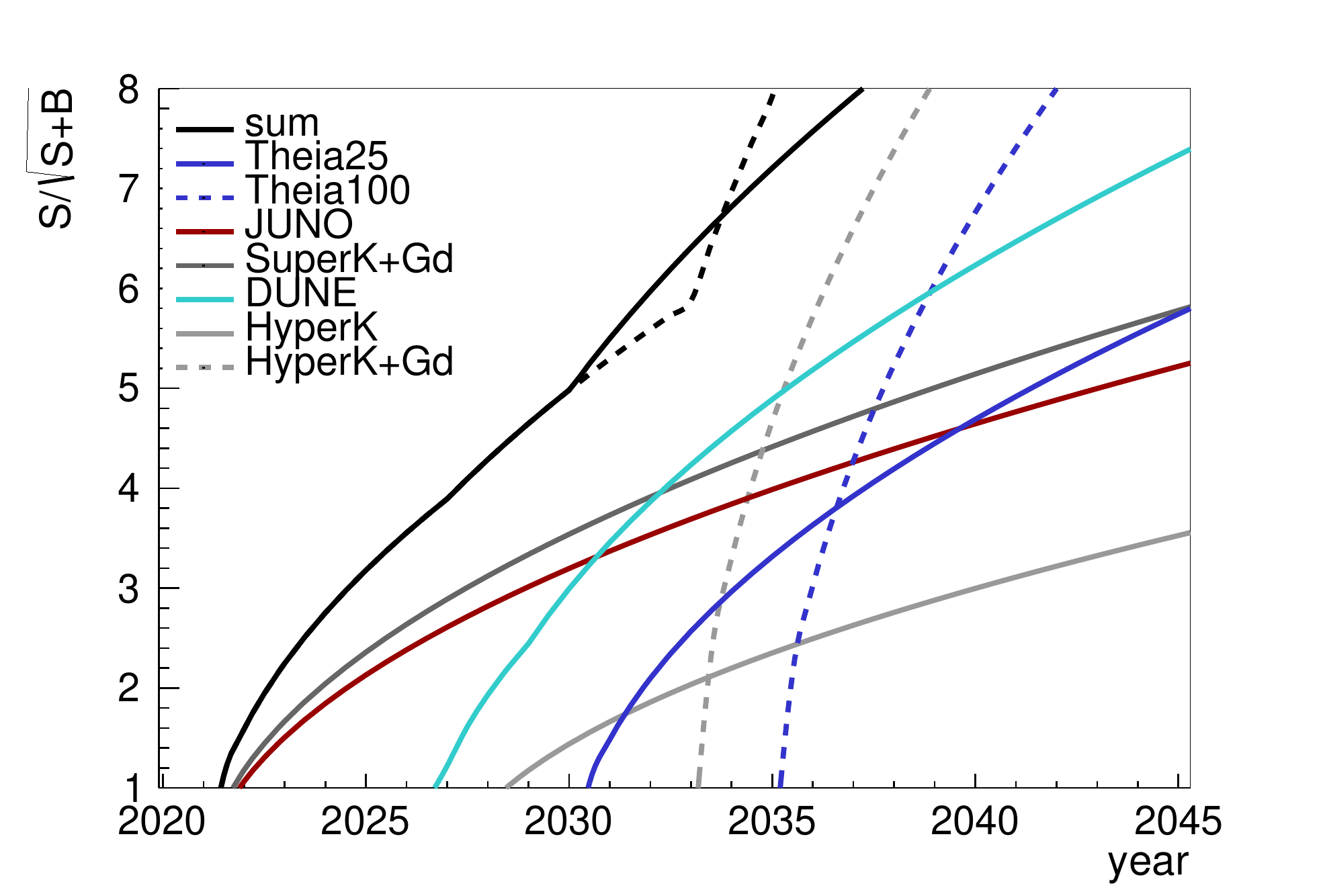}
\caption[]{Projections for the signal rates (left panel) and signal significance (right panel) of the relevant DSNB observatories over the next two decades. Optimistic scenarios correspond to dashed lines. The optimistic sum includes Theia100, and a second tank for Gd-loaded HyperK. DUNE is not added to the overall sum, due to different neutrino channel. Once initiated, Theia100 and HK-Gd soon dominate the scene regarding both collected signal statistics and significance of the detection. Theia25 makes a slower start but provides an increasingly relevant contribution over ten years of data taking. See the text for a more detailed discussion.}
\label{Fig:Comp_Rates}
\end{figure*}
The detection potential for the DSNB in a WbLS detector is best evaluated in comparison to the capabilities of other large-scale neutrino observatories coming online during the next decade. At the time of writing, there is no experiment offering sufficient sensitivity for a positive detection, with the Water Cherenkov Detector (WCD) Super-Kamiokande (SK) providing the current best limit on the DSNB flux. Contrariwise, the currently prepared SK-Gd phase featuring the addition of gadolinium to the water target as well as Hyper-Kamiokande with (HK-Gd) or without gadolinium (HK) all promise genuine sensitivity to the DSNB signal \cite{Abe:2011ts,Abe:2018uyc}. Moreover, JUNO will offer the possibility for DSNB search in an organic liquid scintillator (oLS) \cite{An:2015jdp}, while the DUNE liquid-argon~(LAr) TPCs promise sensitivity to the $\nu_e$ flux component of the DSNB \cite{Cocco:2004ac}.
\medskip\\
{\bf Signal over background.} Tab.\,\ref{Tab:Comp} permits a coarse comparison of these experiments concerning their prospects for a DSNB detection: The experiments are sorted by the underlying detector technology.\ The center columns list the fiducial masses foreseen, the scheduled start of operation, and the observation window. To improve comparability, we used the cross-sections, masses and efficiencies given in the relevant experimental references \cite{Talk_Vagins,Talk_Takatomi,Abe:2018uyc,Abi:2018dnh} and calculated the expected event rates for the fiducial DSNB model \cite{Kresse:2020} used throughout the present paper. Background rates were scaled for exposure. Based on this, we show the number of signal and background rates for a uniform exposure of 100\,kt$\cdot$yrs in the third and second to last column. Finally, the significance of the signal over background, $S/\sqrt{S+B}$, is displayed in the last column.

This comparison illustrates the exceptional performance of WbLS as a target material:
With $\sim$\,17 events detected in 100\,kt$\cdot$yrs, WbLS is leading in signal acceptance, i.e.~the detection efficiency per unit exposure, and thus permits a fast accumulation of statistics.
More importantly, WbLS features also the largest signal significance over background of $\sim$3.4, thereby shortening the exposure required to claim a discovery of the DSNB signal.
\medskip\\
\noindent{\bf Time projections of exposure.} While WbLS shows the best performance per unit exposure, the sensitivity of a given experiment depends effectively on the total target mass it commands and the start of operation. The corresponding projections for the individual experiments are displayed in fig.\,\ref{Fig:Comp_Rates}: The left panel shows the DSNB signal statistics as a function of calendar year, the right panel the corresponding signal significance. Again, we base our projections on the information given in Refs.~\cite{Talk_Vagins,Talk_Takatomi,Abe:2018uyc,Abi:2018dnh}.

In the initial phase, the scene will be dominated by SK-Gd and JUNO since both will start data taking around 2021. While SK-Gd will collect almost twice the statistics per year compared to JUNO, the JUNO S:B-ratio is significantly better, leading to a comparatively small lead of SK-Gd in signal significance. DUNE is estimated to start data taking 2026 with two caverns ($\sim$\,20\,kt$\cdot$yrs), while the third and fourth chamber will be ready one and three year(s) later, respectively \cite{Abi:2018dnh}. Around 2027, HK will start data taking. It is worth noting that despite of the huge target mass and hence fast accumulation of signal statistics, its sensitivity will not surpass that of SK-Gd. However, DSNB signal significance will sky-rocket as soon as gadolinium is added for HK-Gd. DUNE, on the other hand, will reach signal statistics and significance comparable to SK-Gd about ten years into its operation phase (with the usual caveat that this concerns only the $\nu_e$ signal of the DSNB).  

Both realizations of Theia have to be regarded against this background.\ In the following, we assume an optimistic start of operation in 2030 (2035) for Theia25 (Theia100).\ Set on this time scale, Theia100 would immediately emerge as one of the two leading DSNB observatories, providing both fast collection of signal statistics and excellent signal significance. While HK-Gd will collect higher numbers of events, Theia100 profits from the considerably better S:B ratio. Depending on the relative starts of data taking, Theia100 bears the promise of a speedy discovery of the DSNB signal and an important contribution to the follow-up phase of DSNB spectroscopy.

Even the smaller Theia25 will be able to make a relevant contribution. Despite a relatively long ramp-up time ($>10$\,yrs), Theia25 will eventually surpass JUNO in collected event statistics and approach the event numbers of SK-Gd. The same is true for signal significance. Before the arrival of HK-Gd, arguably the most likely scenario for a further exploration of the DSNB is a combined analysis of the data sets of all running experiments (i.e.~SK-Gd and JUNO for the $\bar\nu_e$ component). Especially in this scenario, a contribution of Theia25 would prove very important, in terms of both event statistics and understanding of the crucial NC atmospheric background. While not as spectacular as Theia100, Theia25 would thus provide a substantial improvement of global DSNB sensitivity. \\

% -----Conclusion--------
\section{\label{sec:conclusions}Conclusions}
As laid out in the present paper, WbLS will provide an excellent target material for the detection of the DSNB. We investigated this potential for two possible configurations of the future neutrino observatory Theia. Performing a full analysis including IBD event selection, basic discrimination cuts as well as a selection of events featuring single-rings, high C/S ratio and no delayed decays of final-state radio-nuclei, we find a remaining signal efficiency of $>$80\,\% and a background residual of 1.3\,\% of the high-level selection cuts. Based on a statistical analysis, we conclude that an exposure of $\sim$\,190\,kt$\cdot$yrs will be sufficient to claim a $5\sigma$ discovery of the DSNB under standard assumptions. Longer measuring times may be required in case detector uncertainties are unexpectedly large or the DSNB is best described by a low-flux model~\cite{Kresse:2020}.

While such comparisons are always difficult, we also tried to evaluate the detection potential of a WbLS detector in the context of other large-scale neutrino observatories (present and future) that feature sensitivity for the DSNB. We conclude that per unit detector volume, WbLS outperforms all other detection techniques. However, such an evaluation must take into account as well the time scales on which measurements can be performed and the target masses that can be realized: If realized in the mid-2030s, Theia100 would quickly emerge as a leading DSNB observatory $-$ only the considerably larger HK-Gd will provide similar sensitivity. The more modest Theia25 would take another ten years before having collected sufficient data to add significantly to a global analysis of the DSNB flux and spectrum.

%Acknowledgements
\begin{acknowledgments}
This work was supported by the Collaborative Research Center SFB1258 of the Deutsche Forschungsgemeinschaft (DFG), the Excellence Cluster Universe and Excellence Cluster ORIGINS.
We thank Lothar Oberauer, Gabriel Orebi Gann, and Bob Svoboda for valuable discussions.
DK is grateful to Thomas Ertl and Hans-Thomas Janka for their continued support and guidance during the project.
\end{acknowledgments}
%\nocite{*}

\bibliography{Bibliography}% Produces the bibliography via BibTeX.

\end{document}